\documentclass[12pt,times]{elsarticle}
\usepackage{rotating}
\usepackage{multirow}
\usepackage{tabularx,booktabs}
\usepackage[T1]{fontenc}
\usepackage{amsbsy}
\usepackage{amsfonts}
\usepackage{amsmath}
\usepackage[]{natbib}
\usepackage[usenames,dvipsnames]{color}
\usepackage{epstopdf}
\usepackage{float}
\usepackage{placeins}
\usepackage{enumitem}
\usepackage[x11names]{xcolor}
\usepackage[colorlinks]{hyperref}
\usepackage{graphicx}
\usepackage{geometry}
\AtBeginDocument{\hypersetup{citecolor=blue,urlcolor=brown,linkcolor=Blue}}
\usepackage[font=normalsize,labelfont=bf]{caption}
\usepackage{subcaption}
\usepackage{float}
\DeclareCaptionFormat{cont}{#1#2#3\par}

\journal{International Journal of Plasticity}

\bibpunct{\textcolor{blue}{(}}{\textcolor{blue}{)}}{\textcolor{blue}{;}}{a}{\textcolor{blue}{,}}{\textcolor{blue}{,}}

\biboptions{authoryear}

\begin{document}
	
\begin{frontmatter}
	
\title{Modeling the Effects of Slip, Twinning, and Notch on the Deformation of Single-Crystal Austenitic Manganese Steel}

\author[IPPT]{S.~Virupakshi}
\author[TAMU]{X.~Zheng}
\author[IPPT]{K.~Frydrych}
\author[TAMU]{I.~Karaman}  
\author[TAMU]{A.~Srivastava} 
\author[IPPT]{K.~Kowalczyk-Gajewska\corref{mycorrespondingauthor}}
\ead{kkowalcz@ippt.pan.pl}	
	
\address[IPPT]{Institute of Fundamental Technological Research, Polish Academy of Sciences, Pawi\'nskiego 5B, 02-106 Warsaw, Poland}
\cortext[mycorrespondingauthor]{Corresponding author}

\address[TAMU]{Department of Materials Science \& Engineering, Texas A\&M University, College Station, TX 77843, USA}
	
\begin{abstract}
The objective of this work is to deconvolute the interaction of slip, twinning, and notch on the deformation response of an austenitic manganese (Hadfield) steel using detailed finite element simulations. The simulations employ a rate-dependent crystal plasticity constitutive model that incorporates both slip and twinning deformation mechanisms. The model accounts for the spatially non-uniform appearance of new twin-related orientations, hardening due to slip–twin interactions, and modified properties of the twinned crystal. Limited experiments on single-crystal dog-bone and single-edge notch specimens, with two crystal orientations, are also conducted to aid the simulation. Several features of the experimental observations are accurately captured in the simulations. For example, simulations accurately capture distinct stress–strain responses associated with different crystallographic orientations, including variations in initial hardening behavior followed by either decreasing or increasing hardening depending on the dominant deformation mechanisms. The simulation also captures the observed orientation-dependent asymmetric deformation of the notch in single-edge notch specimens. Additionally, by selectively activating deformation mechanisms, the role of twinning is isolated and its influence on both global and local response is clearly demonstrated. These results provide a mechanistic understanding of how deformation mode interactions and local geometry (i.e., notch) influence the response of these materials.
\end{abstract}

\begin{keyword}
Austenitic steel, Single crystals, Anisotropy, Deformation twinning, Crystal plasticity, Fracture, Finite element method	\end{keyword}
	
\end{frontmatter} 	

\section{Introduction}
\label{Intro}

Austenitic manganese steels, known as Hadfield steels, are characterized by their high strain-hardening rate, exceptional ductility, toughness, and wear resistance, and are widely employed in applications such as railroads, heavy machinery, and armored vehicles \citep{subramanyam1990austenitic, havlicek2012experience, havel2017austenitic, okechukwu2018development}. Their remarkable mechanical behavior is primarily attributed to the activation of deformation twinning, which has been experimentally studied using mechanical testing of both polycrystalline \citep{raghavan1969nature, dastur1981mechanism, adler1986strain, hutchinson2006dislocation, bayraktar2004deformation} and single crystal specimens \citep{Karaman00, karaman2001extrinsic, karaman1998mechanisms, karaman2000deformation, efstathiou2010strain, canadinc2007orientation, zheng2025interaction}. Twinning activity has also been recognized as an important factor influencing deformation and failure in other manganese steels; see for example \citep{Ding24, Shen24}.

Specifically, in the recent in-situ experimental work by \cite{zheng2025interaction}, the deformation and fracture behavior of austenitic manganese steels was studied using both single-crystal dog-bone and single-edge notch tension specimens. These experiments showed that, in austenitic manganese steels, twinning continues to evolve with increasing deformation, and both twinned and untwinned regions undergo significant plastic deformation. Importantly, these experiments showed that, in a single-edge notch specimen oriented for twinning as the dominant deformation mechanism, the initially symmetrical notch undergoes large asymmetrical deformation. With continued deformation, multiple microcracks nucleate at the intersections of the deformed notch and twin boundaries. In contrast, in single-edge notch specimens oriented for crystallographic slip as the dominant deformation mechanism, the notch deforms symmetrically, with a single crack eventually nucleating from the center of the deformed notch.

While these experiments provide valuable insight, to the best of our knowledge, there is currently no work on modeling the effects of slip, twinning, and notch on the deformation of austenitic manganese steels. Existing crystal plasticity finite element (CPFE)-based modeling studies on notched specimens have mainly focused on other single- or polycrystalline metals and have rarely incorporated deformation twinning. For example, \cite{Selvarajou16} investigated the micromechanics of anisotropic plastic flow via combined slip and twinning in single crystals, and \cite{Baweja24} modeled polycrystalline circular-notched specimens of magnesium. Similarly, \cite{Subrahmanya17} performed plane strain simulations of cylindrical void growth ahead of a notch tip in magnesium single crystals. More recently, \cite{Sahu24} examined single V-notch CP-Ti polycrystals both experimentally and through CPFE simulations, but their modeling was restricted to the notch region, which was treated as a second-phase material with very low stiffness, and twinning was neglected. Furthermore, the interaction of twinning and notch in austenitic manganese steels is unlike that in magnesium or its alloys and other quasi-brittle materials, where twin boundaries serve as weak interfaces or discontinuities in the absence of significant plastic deformation, simply providing a pathway for crack growth \citep{kaushik2014experimental, kaushik2014finite}.

Following these, the objective of this work is to deconvolute the effects of slip and twinning on the deformation response of both dog-bone (which induces uniaxial loading) and single-edge notch (which induces more complex multiaxial loading) tension specimens using detailed CPFE simulations. These simulations allow us to selectively introduce deformation mechanisms, enabling a clear evaluation of their individual roles. Furthermore, there is always variability in experiments, such as the orientation of the single-crystal specimens. As such, parametric studies are also carried out to understand the effect of these misorientations. To aid these simulations, limited experiments on both single-crystal dog-bone and single-edge notch tension specimens of Hadfield steels, with axes aligned close to $\langle 001 \rangle$ or $\langle 111 \rangle$, have also been conducted.

The FE simulations carried out herein involve modeling large deformations in three-dimensional dog-bone and single-edge notch tension specimens using a continuum rate-dependent crystal plasticity constitutive model with deformation twinning. For FCC austenitic manganese steels, there are twelve possible slip systems and twelve possible twin systems, all of which are considered in this work. The constitutive model with twinning accounts for three main effects: (i) crystal reorientation \citep{Tome91,Staroselsky98,Kowalczyk13}, (ii) hardening with latent directional effects from slip–twin interactions \citep{Karaman00,Proust07}, and (iii) modified properties for the crystal formed after reorientation \citep{Frydrych20}. A key novelty of the present approach is the detailed treatment of the third aspect.

There are numerical challenges in incorporating twin-related reorientation in the CPFE method. In particular, when twinning criteria are met at a material point, a new crystal orientation must be assigned, which can lead to numerical instabilities. Typically, a mean-field two-phase model by \cite{Kalidindi01} is used until the twin volume fraction reaches a critical value, often around 0.9, at which point the crystal is reoriented and no longer deforms by twinning \citep{Zhang12,LiuWei14}, resulting in smooth stress and strain fields that contradict experimental observations. In this work, we use a probabilistic twin-volume-consistent scheme \citep{Kowalczyk11}, allowing parent and twinned orientations to coexist within the twinning domain and accounting for the volume effect of twins in the simulations. The CPFE implementation, together with the reorientation scheme, is performed using the AceGen package---a tool of automatic differentiation enabling us to obtain the consistent tangent operator and computationally efficient code \citep{Korelc02}. Moreover, a few techniques recently proposed in the literature \citep{Dittmann23,Manik22,Korelc14} have been combined to further enhance the computational efficiency of the framework.

We note that there are several approaches to modeling of deformation twinning, such as the phase-field method, which not only accounts for the nucleation and evolution of twins but also their morphology \citep{RezaeeHajidehi22, Mo22, LiuRotersRaabe23}. Additional approaches include incorporating the twin evolution law into the constitutive description \citep{Qiao16, Cheng18}, using post-remeshing tools to explicitly introduce twins within the grains \citep{Ardeljan15}, or modifying the single-crystal energy function by adding an interaction energy term to penalize the simultaneous existence of the matrix and the twin \citep{Chester16, Huang22}. However, these methods require knowledge of additional parameters and are extremely computationally expensive. As such, modeling the full three-dimensional specimen with all possible slip and twin systems---which is necessary to simulate the response of a single-edge notch specimen---is practically unrealistic. There are also formulations that account for twinning–detwinning effects \citep{Wang13}, which have been extended and applied in the context of the mean-field elastic–viscoplastic self-consistent model \citep{Qiao17,Cheng24} and implemented in the FE method \citep{Yaghoobi22}. Temperature effects have also been incorporated \citep{Hollenweger22} into the variational CP framework \citep{Chang15}. However, since cyclic loading and temperature effects are not the focus of our work, the complexity of the proposed approach is considered sufficient in the present context, leading to a computationally efficient numerical implementation.

Our results show that the CPFE simulations using a fully calibrated constitutive model accurately capture the deformation response of both $\langle 001 \rangle$-oriented (which preferentially activates slip) and $\langle 111 \rangle$-oriented (which preferentially activates twinning) single-crystal dog-bone specimens of austenitic manganese steel. The simulations also capture the relative differences in the response of $\langle 001 \rangle$- and $\langle 111 \rangle$-oriented single-edge notch specimens, as well as accurately predict the differences in the local deformation of the notches in these specimens. Furthermore, by selectively activating slip and twinning in the simulations, we clearly evaluate their individual roles. We also assess the impact of differences in the orientation of the single-crystal specimens on the interaction of slip, twinning, and the notch in the deformation behavior of single-crystal austenitic manganese steel. The remainder of the paper is organized as follows: a brief description of the experiments is provided in Section 2, the numerical method is detailed in Section 3, the results and discussion are presented in Section 4, and the conclusions are drawn in Section 5.

\section{Experimental method}
\label{Sec:2}

The material used in this work is single crystals of Hadfield steel with a nominal composition of 13 wt\% manganese, 1.1 wt\% carbon, and the remainder being iron, as in \cite{zheng2025interaction} and other experimental works \citep{ karaman1998mechanisms, Karaman00, karaman2000deformation, karaman2001extrinsic}. The single crystals were grown using the Bridgman technique in a helium atmosphere, with the crystal growth direction approximately aligned along either the $\langle 001 \rangle$ or $\langle 111 \rangle$ crystallographic orientations. These crystals were then subjected to homogenization heat treatment at 1373 K for 24 hours in an inert gas atmosphere. Subsequently, flat dog-bone specimens with a gauge length of approximately 8 mm, a width of approximately 3 mm, and a thickness of approximately 0.7 mm, aligned along the crystal growth direction, were machined using wire electrical discharge machining. After machining, the specimens were solutionized at 1373 K for 1 hour and then quenched in water at room temperature. {{All solutionized single-crystal specimens exhibited single-phase austenite without retained carbides or second phases \citep{zheng2025interaction, karaman1998mechanisms, Karaman00, karaman2000deformation, karaman2001extrinsic}, consistent with broader studies on high-Mn austenitic steels showing that solution annealing fully dissolves carbides in alloys with carbon contents near 1 wt\% (as in Hadfield steel), yielding a homogeneous single-phase austenite \citep{gurol2020effect}.}}

In a subset of flat dog-bone specimens, a notch approximately 1.5 mm long with a root radius of approximately 45 $\mu$m was machined across the gauge width in the middle of the gauge section using a diamond wire saw. All the mechanical tests were performed using a miniature tension/compression module (Kammrath \& Weiss) under a high-resolution digital optical microscope (Olympus DSX510), which allowed us to capture images of the deforming sample. This was particularly important for capturing the local deformation of the notch in single-edge notch specimens.

{{Under the monotonic room-temperature loading considered here, long-range diffusion of Mn and C is minimal; accordingly, we assume a chemically homogeneous austenite and do not consider solute segregation to interfaces. This is consistent with observations in Fe–Mn–C austenitic steels that deformation-induced (coherent) twin boundaries exhibit negligible segregation compared with non-coherent/high-angle grain boundaries \citep{herbig2015grain}.}}

In \cite{zheng2025interaction}, where the primary focus was on $\langle 111 \rangle$-oriented single crystals of Hadfield steel, extensive experimental analysis of deformation-induced microstructure evolution was carried out using ex-situ and in-situ scanning electron microscopy (SEM) and electron backscattered diffraction (EBSD) analysis. Additionally, fractographic analysis of the fracture specimens was conducted using SEM imaging. Detailed analysis of twin formation and their structure in Hadfield steel has also been carried out extensively using transmission electron microscopy (TEM) in \citep{ karaman1998mechanisms, Karaman00, karaman2000deformation, karaman2001extrinsic}. These results are discussed in the context of the current modeling results wherever relevant.

\section{Numerical method}

The effects of crystallographic slip, twinning, and the presence of a notch in single-crystal austenitic manganese steel are analyzed using three-dimensional finite deformation FE simulations of dog-bone and single-edge notch tension specimens. These simulations employ a continuum rate-dependent crystal plasticity constitutive model that accounts for deformation twinning. The crystal plasticity constitutive model and the FE implementation are described below.

\subsection{Crystal plasticity constitutive model}
\label{Sec:CP}

The crystal plasticity constitutive model used to carry out FE simulations of FCC single crystals deforming by slip and twinning follows from the work of \cite{Kowalczyk10}. However, in that prior work, the model was formulated in a velocity-based framework, neglecting elastic strains, and was combined with mean-field models of polycrystals: the iso-strain Taylor approach and the VPSC model \citep{Lebensohn93}. In this work, the model has been adapted for incorporation into CPFE method.

We consider the elastic-viscoplastic model of a single crystal formulated in a displacement-based framework, and the kinematic description follows standard formulations \citep{HILL1972401, Asaro85}. The deformation gradient $\mathbf{F}$ is multiplicatively decomposed into two parts:

	\begin{equation}  \label{eq:Multiplicative}
		\mathbf{F}=\mathbf{F}_e\mathbf{F}_p\
	\end{equation}
    where $\mathbf{F}_e$ and $\mathbf{F}_p$ denote the elastic and plastic components, respectively. The evolution of the plastic part of the deformation gradient is governed by the equations:
	\begin{equation}\label{Eq:Fpevolution}
		\dot{\mathbf{F}}_p=\hat{\mathbf{L}}_p\mathbf{F}_p,
	\end{equation}
	where the dot over the quantity denotes its material time derivative.
	
The usual approach \citep{Chin69,Kalidindi98,Kalidindi01,Staroselsky98,Staroselsky03} is used to account for twinning in the crystal plasticity model. Twinning is considered as a pseudo unidirectional slip mode. The rate of pseudo slip $\dot{\gamma}^l_{(t)}$ is related to the rate of volume fraction $\dot{f}^l$ of the twinned part formed by the twin system $l$ as
\begin{equation}\label{eq:twinvolevolution}
    \dot{\gamma}^l_{(t)} = \gamma^{TW} \dot{f}^l,
\end{equation}
where $\gamma^{TW}$($=1/\sqrt{2}$ for FCC crystal) is the characteristic twin shear strain. Denoting by $\dot{\gamma}^k_{(s)}$ as the rate of slip on the $k-th$ slip system, the plastic part of the velocity gradient $\hat{\mathbf{L}}_p$ is described as follows:

\begin{equation}
\hat{\mathbf{L}}_p=\sum_{k=1}^{2M}\dot{\gamma}^k_{(s)}\mathbf{m}^k_0\otimes\mathbf{n}^k_0
+\sum_{l=1}^{N}\dot{\gamma}^l_{(t)}\mathbf{m}^l_0\otimes\mathbf{n}^l_0 = \sum_{r=1}^{2M+N}\dot{\gamma}^r\mathbf{m}^r_0\otimes\mathbf{n}^r_0,
\label{velgrad}
\end{equation}
where $\mathbf{m}^r_0$ and $\mathbf{n}^r_0$ are the unit vectors along the slip/twin direction and normal to the slip/twin plane of the system $r$ in the reference configuration, $M$ denotes the number of slip systems and $N$ the number of twinning systems, respectively; see Table \ref{tab:slipandtwinsystems}.

\begin{table}[t]
    \caption{Slip and twin systems in FCC austenitic manganese steel considered in this work.}
    \label{tab:slipandtwinsystems}
    \centering
    \begin{tabular}{|l|l|l|l|}
    \hline
    & Slip/Twin plane & Slip/Twin direction & Number of systems \\
    \hline
   Slip & \{111\} & $\langle\bar{1}10\rangle$ & 12\\
    Twinning & $\{ 111 \}$ & $\langle2{1}1\rangle$ & 12\\
    \hline
    \end{tabular}
\end{table}

To unify the description of slip and twinning in Eq.~(\ref{velgrad}), the slip in the \textbf{m} direction is distinguished from the slip in the $\mathbf{\bar{m}}$ direction, and $\dot{\gamma}^r$$(\geq 0)$ denotes the magnitude of shear rate in the $r$-th deformation mode.  

The rate of shearing on a given slip or twinning system $r$ is calculated using the visco-plastic power law \citep{Asaro85}:
\begin{equation}
\dot{\gamma}^r=\dot{\gamma}_0 \,
\left(\frac{\tau^r}{\tau_c^r}\right)^n\,,
\end{equation}
\noindent where $\dot{\gamma}_0$ is the reference shear rate, $n$ is the strain-rate-sensitivity exponent, $\tau^r$ is a non-negative resolved shear stress and $\tau_c^r$ its critical value. The resolved shear stress is defined by the projection of the Mandel stress tensor $\mathbf{M}_e$ on the plane of slip $\mathbf{n}^r_0$ and in the direction $\mathbf{m}^r_0$:
	\begin{equation}
	\tau^r=\langle\mathbf{m}^r_0\cdot\mathbf{M}_e\cdot\mathbf{n}^r_0\rangle \geq 0,\hspace{0.2cm}
	\mathbf{M}_e = \mathbf{F}^T_e \mathbf{S} \mathbf{F}^T_p = \mathbf{F}^T_e \boldsymbol{\tau} \mathbf{F}^{-T}_e \,,
\end{equation}
where $\langle\cdot \rangle\equiv\frac{1}{2}((\cdot)+|\cdot|)$, $\mathbf{S}$ is the first Piola-Kirchhoff stress and $\boldsymbol{\tau}$ is the Kirchhoff stress. The Mandel stress tensor is obtained using a hyper-elastic law:
\begin{equation}
	\mathbf{M}_e=2\mathbf{C}_e\frac{\partial \Psi}{\partial \mathbf{C}_e}\,,
\end{equation}
where $\mathbf{C}_e=\mathbf{F}^T_e\mathbf{F}_e$ is the right elastic Cauchy-Green tensor and
\begin{equation}
	\Psi=\frac{1}{2}\mathbf{E}_e\cdot\mathbb{L}^e\cdot\mathbf{E}_e
\end{equation}
is the  {Kirchhoff-type function of} free energy density per unit volume in the reference configuration, $\mathbb{L}^e$ is the anisotropic stiffness tensor of single crystal and $\mathbf{E}_e=\frac{1}{2}(\mathbf{C}_e-\mathbf{1})$ is the elastic Lagrangian strain tensor.

The critical resolved shear stress $\tau_c^r$ evolves differently with accumulated slip and accumulated twin volume fraction according to the hardening laws \citep{Kowalczyk10,Kowalczyk11} developed based on prior experimental observations \citep{Kalidindi01,Szczerba04,Karaman00,Karaman01,Sahoo19,Sahoo20}. 

Hardening of slip systems due to slip and twinning activity is described as follows:
\begin{equation}\label{SlipHard}
\dot{\tau}_c^r=\dot{\tau}_c^{r+M}=H_{(ss)}^{r}\sum_{q=1}^{M}h^{(ss)}_{rq}\dot{\bar{\gamma}}^q+H_{(st)}^{r}
\sum_{q=2M+1}^{2M+N}h^{(st)}_{rq}\dot{\gamma}^q\,\quad (r\leq M)\,,
\end{equation}
where $\dot{\bar{\gamma}}^q=\dot{\gamma}^q+\dot{\gamma}^{q+M}$. The latent hardening on a given system $r$ due to activity on system $q$ is described using the submatrices $h_{rq}^{(\alpha\beta)}=q^{(\alpha\beta)}+(1-q^{(\alpha\beta)})|\mathbf{n}^r\cdot\mathbf{n}^q|$, where $q^{(\alpha\beta)}$ ($\alpha,\beta=t,s$) are the latent hardening parameters and the term $|\mathbf{n}^r\cdot\mathbf{n}^q|$ is responsible for considering the influence of coplanarity of systems.
$H_{(ss)}^{r}$ and $H_{(st)}^{r}$ denote hardening moduli of slip systems due to slip and twinning activity,
\begin{equation}\label{Hss}
    H_{(ss)}^r=h_1^{ss}+ h_0^{ss}\exp{\left(\frac{-h_0^{ss}}{\tau_{sat}^{ss}} \Gamma^s\right)}\;, \quad 
    H_{(st)}^{r}= h_0^{st}\left(\frac{f^{\rm{TW}}}{f^{st}_{sat}-f^{\rm{TW}}}\right)
\end{equation}
with
\begin{equation}\label{eq:11}
    	\Gamma^s=\int \dot{\Gamma} {\rm{d}}t,
		\hspace{0.2cm}
		\dot{\Gamma}=\sum_{r=1}^M{\dot{\bar{\gamma}}^r}, 
\end{equation}
where $h_0^{ss},h_1^{ss}, h_0^{st}, \tau_{sat}^{ss}$ and $f_{sat}^{st}$ are hardening parameters and $f^{\rm{TW}}$ is the total volume fraction of twins. 

	On the other hand, hardening of twin systems due to slip and twinning activity is specified as follows:
\begin{equation}\label{twh1}
\dot{\tau}_c^r=H_{(ts)}^{r}\sum_{q=1}^{M}h^{(ts)}_{rq}\dot{\bar{\gamma}}^q+H_{(tt)}^{r}
\sum_{q=2M+1}^{2M+N}h^{(tt)}_{rq}\dot{\gamma}^q\, \quad (r >2M)\,.
\end{equation}
$H_{(ts)}^{r}$ and $H_{(tt)}^{r}$ denote hardening moduli of twin systems due to slip and twinning activity and are specified as
\begin{equation}\label{Hst}
    H_{(ts)}^r=h_1^{ts}+ h_0^{ts}\exp{\left(\frac{-h_0^{ts}}{\tau_{sat}^{ts}} \Gamma^s\right)}, \quad 
    H_{(tt)}^{r}= \frac{h_0^{tt}}{\gamma^{\rm{TW}}}\left(\frac{f^{\rm{TW}}}{f^{tt}_{sat}-f^{\rm{TW}}}\right)
\end{equation}
Similarly, as in Eq.~\eqref{Hss}, $h_0^{ts},h_1^{ts}, h_0^{tt}, \tau_{sat}^{ts}$ and $f_{sat}^{tt}$ are hardening parameters to be identified for the material of interest.

In the present study, in order to reduce the number of parameters, hardening of slip due to twinning is neglected ($H^{r}_{(st)}=0$) and the latent hardening coefficient $q^{(ss)}=1$. Moreover, hardening of twin due to slip is specified solely by the linear term in Eq.~(\ref{Hst})$_1$, so that $h_0^{(ts)}=0$ ($\tau_{sat}^{ts}$ is then redundant), and the latent hardening coefficients $q^{(ts)}=q^{(tt)}=1$. Some remarks on the influence of the $q^{(\alpha\beta)}$ value are included in \ref{app2}. An extensive discussion on the physical nature of latent hardening for slip–twin interactions can be found in \cite{ElKadiri10}.

In the event that the reorientation condition is true for a given point, and the crystal lattice at this point has been reoriented due to twinning, the newly formed crystal is assumed to be described by the same material model, but with different material parameters. This accounts for the fact that the mechanical characteristics of twins differ from those of the parent crystal mainly due to Basinski’s mechanism, i.e., the change of glissile dislocations into sessile ones \citep{Sahoo19}. Secondary twinning is more difficult to initiate than slip in twinned regions. This is because twinning evolution in Hadfield steel is usually observed in the form of multiple thin layers, which eventually join rather \citep{zheng2025interaction} than the growth of a single twin as in magnesium. Thus, in the present calculations, we assumed that twinning is no longer possible in reoriented regions, while the initial critical shear stress $\tau_{c}^{0}$ for slip and the hardening parameters $h_0^{ss}$, $h_1^{ss}$, and $\tau_{sat}^{ss}$ are modified.

The probabilistic twin volume consistent (PTVC) reorientation scheme \citep{Kowalczyk10,Kowalczyk13} is applied to account for lattice reorientation due to twinning. It ensures that the twin volume fraction and the proportion of reoriented domains in the sample are consistent. When the reorientation criterion is met, a crystal lattice is reoriented to twin-related orientation. It should be emphasized that information accessible at the given material point (i.e., Gauss point in FE mesh) is sufficient to validate this reorientation criterion and that at the current deformation step a one-phase material (single orientation) is assumed to exist at this point.
The condition is as follows:
\begin{displaymath}
\textrm{if}\quad (\Delta f^{TW}>\xi\quad \textrm{and} \quad f^{TW}>f_0^{crit}) \quad \textrm{or} \quad f^{TW}>f^{crit} \quad \rightarrow \quad \mathbf{a}^{new}=\mathbf{R}^{TW}\mathbf{a}^{old}
\end{displaymath}
where $\Delta f^{TW}$ is the increment of twin volume for a given time step, $\xi$ is a number randomly generated from the set $(0,\Psi_t)$ and $\Psi_t=1-(f^{TW}-\Delta f^{TW})$, while $f^{crit}$ is a value of twin volume fraction for which the reorientation is always performed. The condition is checked once the volume fraction of twin exceeds $f_0^{crit}$.  $\mathbf{R}^{TW}$ is the rotation tensor
\begin{equation}\label{Eq:RTW}
\mathbf{R}^{TW}=2\mathbf{n}^{TW}\otimes\mathbf{n}^{TW}-\mathbf{I}
\end{equation}
where $\mathbf{n}^{TW}$ is a unit normal to the twin plane of the twin system $l$ for which $\gamma^l=\gamma^{TW}f^l$ is maximum at the given time step. $\mathbf{R}^{TW}$ rotates any crystallographic direction $\mathbf{a}^{old}$ of the parent orientation to the corresponding $\mathbf{a}^{new}$ of the twin-related one. It must be emphasized that the condition is probabilistic in its nature, so, in general, formation of the twin-matrix laminate or distinct twin lamellas is not observed in the calculations. However, parent and twinned orientations co-exist within the FE mesh within the domains of twinning activity, enabling one to account for the volume effect of twins in the simulations.

The present crystal plasticity formulation does not incorporate size effects. Such effects could be included by introducing length-scale parameters, for example, through a size-dependent initial critical stress \citep{Meyers01,Wong16} or by making certain hardening parameters proportional to the mean free path for dislocations or the characteristic width of twin platelets \citep{Karaman01}. Since samples of the same material but with significantly different dimensions were not studied, this extension is not relevant in the present context.

In the current formulation, twinning and detwinning are disabled after reorientation at a given Gauss point to improve computational efficiency. This assumption is not expected to affect the results, as the loading processes considered are nominally proportional, and strain or stress reversal---required for detwinning \citep{Wang13,Yaghoobi20,Frydrych21}---is unlikely. Nevertheless, the model and the PTVC reorientation scheme can account for detwinning or multiple twinning–detwinning events if the post-reorientation grain description allows twinning. In that case, the reorientation scheme operates as for the parent grain, with the rotation tensor in Eq. \eqref{Eq:RTW} defined by the unit normal to the twin plane corresponding to the system with the maximum accumulated twin volume fraction since the last reorientation. The hardening moduli $H_{(st)}$ and $H_{(tt)}$ in Eqs. \eqref{Hss} and \eqref{Hst} are then determined from the total twin volume fraction accumulated over the same period.

{{The constitutive model also does not track solute redistribution or segregation to twin boundaries. This simplification is appropriate for the present solutionized single-crystal Hadfield steel tested under monotonic loading (see Section~\ref{Sec:2}), where deformation-induced twins are expected to show little or no segregation \citep{herbig2015grain}. However, in other loading histories or environments (e.g., hydrogen charging) or in other material systems, segregation at twin boundaries can occur and may influence the deformation mechanisms. Extending the framework to include chemo–mechanical coupling represents an interesting direction for future work.}}

\subsection{Finite element model}

\begin{figure}[!t]
	\begin{tabular}{ll}
		\includegraphics[angle=0,width=0.5\textwidth]{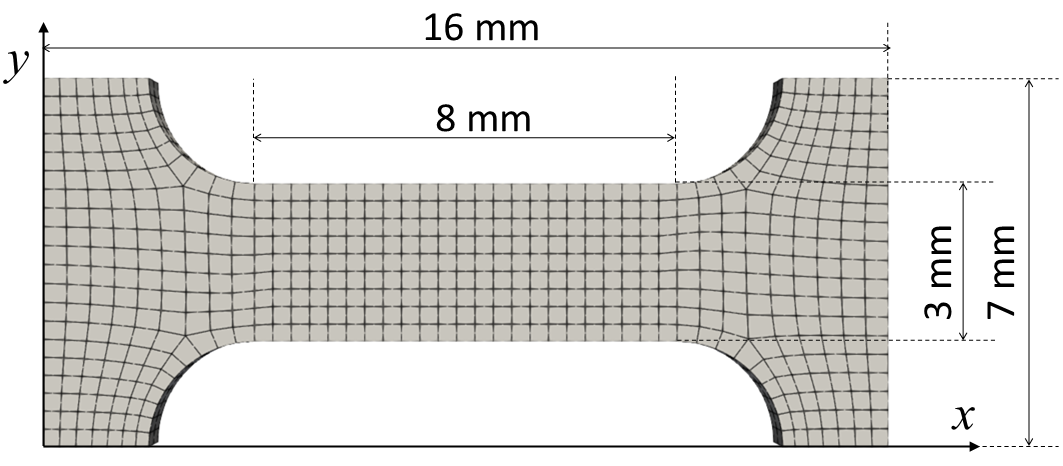}&
		\includegraphics[angle=0,width=0.35\textwidth]{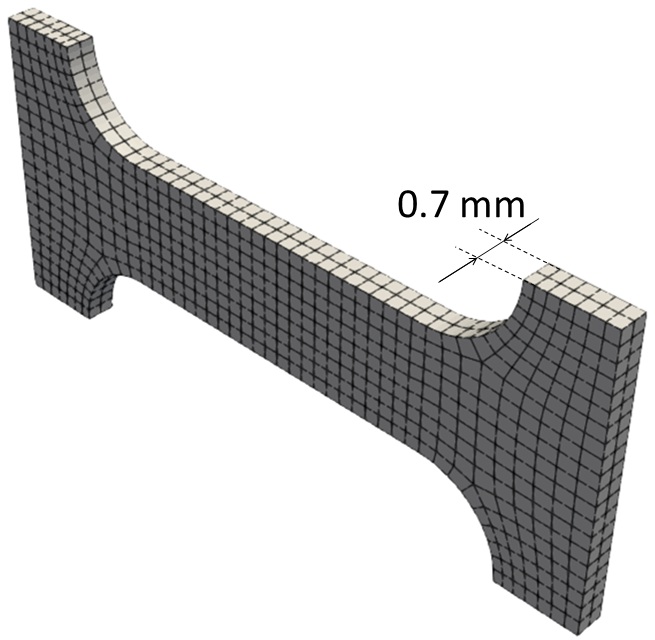}\\
		(a)& (b)\\
		\includegraphics[angle=0,width=0.52\textwidth]{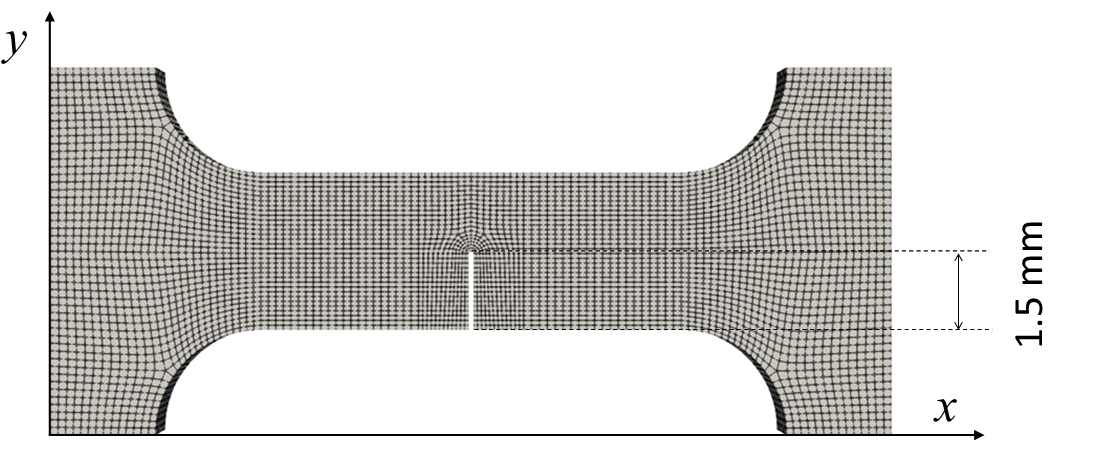}&
		\includegraphics[angle=0,width=0.35\textwidth]{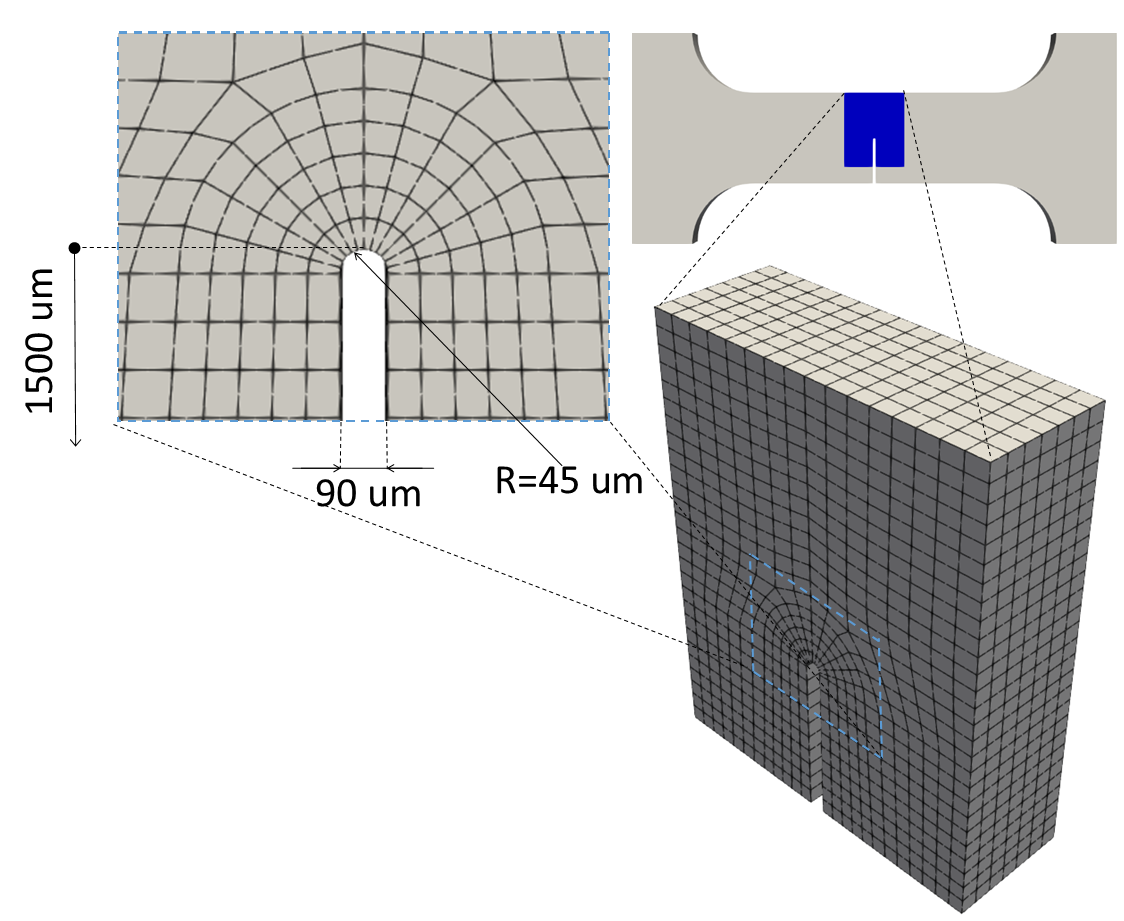}\\
		(c)& (d)
	\end{tabular}
	\caption{Geometry and FE mesh of the specimens: (a)–(b) flat dog-bone and (c) single-edge notch with (d) zoomed-in view of the notch region.}
	\label{fig:schematicview}
\end{figure} 

The full geometry of both dog-bone and single-edge notch specimens, with the same dimensions as in the experiments (Section~\ref{Sec:2}), is modeled for CPFE simulations. The modeled geometry of these specimens along with the FE mesh is shown in Fig.~\ref{fig:schematicview}. Modeling the full geometry of the specimen is essential in this case to account for the boundary conditions imposed in the experiments and to avoid over- or under-constraining the material deformation. The FE mesh for all specimens employs 20-noded brick elements. Since the present CPFE model does not include any intrinsic size effects, the simulated response is invariant to proportional scaling of the sample geometry. The absolute dimensions used in the simulations are therefore chosen to match the single-crystal specimens tested experimentally, and the material parameters are calibrated using those data. Consequently, the calibrated parameters are valid for specimens of similar dimensions and loading conditions.

The FE discretization of the dog-bone and single-edge notch specimens differs, as the presence of a notch is expected to lead to localized deformation in its vicinity. Accordingly, an approximately 2.5-times denser mesh is used for single-edge notch specimens compared to dog-bone specimens, with additional refinement around the notch area. The mesh densities for both specimen types are determined based on a standard mesh sensitivity study. Selected results from this study, related to the robustness of the PTVC scheme, are included in \ref{app1}.

\begin{figure}
	\centering
	\includegraphics[angle=0,width=0.7\textwidth]{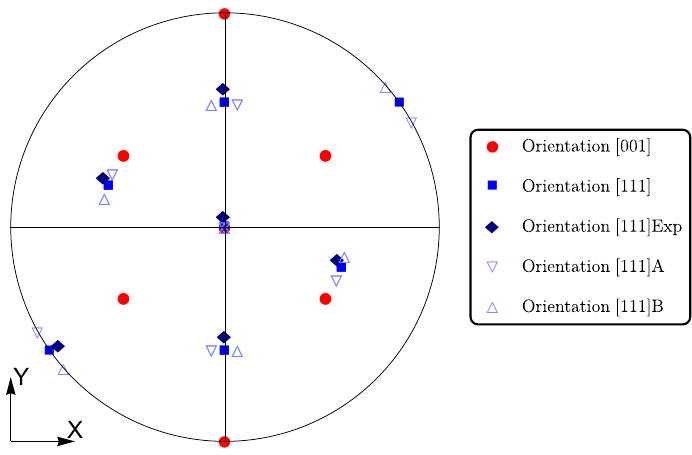}
	\caption{Pole figure $\{011\}$ showing the orientations listed in Table~\ref{tab:crystal orientations}.}
	\label{fig:PF}
\end{figure}

For both specimen types, two main sets of crystal orientations are considered: one where the specimen axis is aligned with the $[001]$ crystallographic direction, and another where it is aligned with or close to $[111]$. For the $[111]$ orientation, parametric studies are also carried out to understand the effect of deviations from the exact $[111]$ orientation. All orientation instances considered here are listed in Table~\ref{tab:crystal orientations}. Orientation $[111]$Exp corresponds to the one measured in the experiment; see also \citep{zheng2025interaction}. The orientation is misoriented with respect to the ideal orientation $[111]$, both in terms of the loading axis and the sample plane. The loading axis is slightly misoriented with the $[\bar{1}\bar{1}1]$ direction (symmetrically equivalent to $[111]$), while the sample plane is not perfectly aligned with the crystal plane $[\bar{1}0\bar{1}]$ (symmetrically equivalent to $[1\bar{1}0]$). This means that the sample plane is not a symmetry plane of the crystal. The total misorientation angle is approximately $6^\circ$. Orientations $[111]$A and $[111]$B are obtained by rotating the ideal $[111]$ orientation by $6^\circ$ clockwise and counterclockwise around the $z$-axis, respectively, so that the sample plane remains a symmetry plane of the crystal. All five orientations are also displayed in the pole figure $\{011\}$ in Fig.~\ref{fig:PF}.

\begin{table}[!htp]
	\caption{Crystal orientation instances considered in simulations with the respective global coordinate axes. Note that the loading directions for orientation $[111]$Exp, and directions $x$ and $y$ for $[111]$A and $[111]$B, are given by the coordinates of the respective sample basis vectors in the crystal frame, while Miller indices are used in the remaining cases.}
	\label{tab:crystal orientations}
	\centering
	\footnotesize{
	\begin{tabular}{|l|c|c|c|l|}
		\hline
		\multirow{2}{4em}{Case} & Primary& Lateral & Lateral & Euler angles\\
		& loading direction ($x$)&  direction ($y$)&direction ($z$)&\\
		\hline
		Orientation $[001]$ &$[001]$&$[110]$ &$[\bar{1}10]$&$\{225^o,90^o,90^o\}$\\
		Orientation $[111]$   & $[111]$ & $[11\bar{2}]$& $[\bar{1}10]$&$\{225^o,90^o,144.74^o\}$\\
		Orientation $[111]$Exp  & $[-0.56,-0.59,0.59]
		$ & $[-0.48,0.81,0.33]
		$ & $[-0.67,-0.096,-0.73]
		$ & $\{278.16^o, 137.36^o,60.47^o\}$
		\\
				Orientation $[111]$A  & $[0.62,0.62,0.49]
				$ & $[0.34,0.34,-0.87]
				$ & $[\bar{1}10]$& $\{225^o,90^o,144.74^o+6.0^o\}$\\
					Orientation $[111]$B  & $[0.53,0.53,0.66]$
					  & $[0.47,0.47,-0.75]
					  $ & $[\bar{1}10]$& $\{225^o,90^o,144.74^o-6.0^o\}$\\
		\hline
	\end{tabular}}
\end{table}

To replicate the boundary and loading conditions of the experiment, the following conditions are imposed in all FE calculations:
\begin{equation} 
\mathbf{u}(0,y,z) = \mathbf{0} \quad \text{and} \quad \mathbf{u}(x_{\text{max}}, y, z) = \dot{\delta} t \mathbf{i}_x 
\end{equation}
where $\dot{\delta}$ is a constant displacement rate equal to $0.01$ mm/s.

\subsection{Finite element implementation}
\label{2.2}

The standard procedures developed for the finite element implementation of finite deformation rate-dependent elasto-plasticity in a fully Lagrangian, displacement-based setting by \cite{Simo98} are employed. Incremental constitutive equations are obtained using an implicit backward-Euler time integration scheme, and Eq.~(\ref{Eq:Fpevolution}) is integrated using the exponential map:

\begin{equation}\label{Eq:map}
	\mathbf{F}_p(t+\Delta t)=\exp(\Delta t\hat{\mathbf{L}}_p)\mathbf{F}_p(t).
\end{equation}

The FE implementation is carried out using the AceGen/AceFEM packages \citep{Korelc02}, following similar implementations for FCC crystals with slip-only mechanisms \citep{Kucharski14, Frydrych18}. The AceGen package facilitates the derivation of an algorithmic consistent tangent, enabling quadratic convergence. The tool combines the symbolic algebra capabilities of Wolfram Mathematica with automatic differentiation and advanced expression optimization techniques. Computations are performed using the associated AceFEM package. The finite elements used are 20-noded hexahedral elements with reduced integration. Regarding the reorientation condition, it is checked at each Gauss point after global convergence is reached. This treatment is, to some extent, similar to that applied by \cite{Zhang12} and \cite{LiuWei14}.

Several techniques recently proposed in the literature have been incorporated to further enhance the computational efficiency of the framework, namely:
\begin{itemize}
\item Reduction of the number of internal variables for which residual equations are solved at the Gauss point level from 24 to 11, following \cite{Dittmann23}. In this approach, 9 components of Mandel stress, accumulated slip, and accumulated twin are used instead of 24 accumulated shears in 24 slip and twin modes (possible when latent hardening coefficients for system interactions are set to 1).
\item Use of the line-search method \citep{Manik22} to track the solution path, which accelerates convergence in the presence of reorientation events.
\item Use of the exact matrix exponent formula \citep{Korelc14} in the exponential map step specified by Eq.~\eqref{Eq:map}.
\end{itemize}

\subsection{Identification of the constitutive parameters}
\label{idpara}

The elastic constants describing the anisotropic elasticity of a single crystal with cubic symmetry for Hadfield steel are assumed as given in Table~\ref{tab:Material parameters1}, following \cite{Clausen98}. Since quasi-static processes are analyzed and strain rate effects are not investigated in the present study, the rate sensitivity parameters of the model, $n$ and $\dot{\gamma}_0$, are pre-set to 20 and 0.001~[1/s], respectively. With these assumptions and the preliminary simplifications regarding interactions between deformation modes discussed in Section~\ref{Sec:CP}, twelve constitutive parameters remain to be identified.

These parameters are determined based on uniaxial tensile test results of specimens without notches, for two single-crystal orientations. For the $\langle 001 \rangle$ orientation, where only slip was experimentally observed, the test results are used to calibrate parameters governing slip–slip interactions in the virgin crystal. For the $\langle 111 \rangle$ orientation, where twinning was the dominant mechanism, the test results are used to identify parameters describing twin–twin and twin–slip interactions in the virgin crystal, as well as those relevant for the reoriented material that deforms by slip alone.

An initial range of possible parameter values was established, and then optimized using a genetic algorithm, following the methodology described in \cite{Frydrych20}. The final set of identified parameters is listed in Table~\ref{tab:Material parameters2}. All calculations in this paper are performed using the same set of material parameters.

\begin{table}[!htp]
	\centering
	\caption{Elastic constants \citep{Clausen98} and other pre-set parameters for single-crystal Hadfield steel. \label{tab:Material parameters1}}
	\begin{tabular}{|c|c|c|c|c|c|}
		\hline
		$C_{11}$ & $C_{12}$  & $C_{44}$ & $\gamma^{TW}$  & $n$  &  $\dot{\gamma}_0$\\
		
		[GPa]&[GPa]&[GPa]&&&\\
		\hline
		204.6&137.7&126.2&$1/\sqrt{2}$&20&0.001\\
		\hline
	\end{tabular}   
\end{table}

\begin{table}[!htp]
	\centering
	\caption{Constitutive parameters identified based on tensile test results of dog-bone specimens. \label{tab:Material parameters2}	}
	\begin{tabular}{|l|c|c|c|c|c|c|}
		\hline
		&  & $\tau_{0}^c$ & $h_0$ & $h_1$ & $\tau_{sat}/f_{sat}$ &  $q$  \\
		System & Interaction & [MPa] & [MPa] &[MPa] & [MPa]/--  &  \\
		\hline
		\multicolumn{7}{|c|}{virgin crystal (slip and twinning active)}\\ \hline
		slip & slip-slip & 126.5 & 6.07 & 379.5 & 48.0 &  1.0   \\
		& slip-twin & --- & --- & --- & --- &  ---    \\
		twin & twin-slip & --- & 0 & 210.0 & --- &  1.0\\
		& twin-twin  & 112.9 & 26.25 & --- & 0.99 &   1.0      \\
		\hline
		\multicolumn{7}{|c|}{crystal after reorientation (only slip active)}\\ \hline
		slip & slip-slip & 139.1 & 154.5 & 49.36 & -39.48 &  1.0   \\ \hline
	\end{tabular}
\end{table}

A few previous studies have attempted to identify CP parameters for single crystals of Hadfield steel. \cite{Karaman00} performed identification based on tensile test results for the $\langle 111 \rangle$ orientation only. Although some similarities exist between their formulation and the present one, the crystal plasticity model---particularly in its hardening description---differs from the current approach, making direct comparison of all parameters not possible. In their study, the initial critical shear stress for slip was slightly smaller than for twinning (110 MPa vs. 115 MPa). A similar qualitative relation between these parameters was also applied in \cite{Kowalczyk10} in the context of the VPSC implementation of the PTVC scheme. In the present study, using experimental curves for two orientations, it was found that both quantities are still close to each other, with the initial $\tau_c$ for twinning smaller than for slip. Quantitatively, these values are of comparable magnitude. In \cite{Lindroos18}, the initial $\tau_c$ was identified as almost equal for slip and twinning mechanisms (87 MPa and 90 MPa, respectively). Material parameters in that work were identified based on the response of two orientations, $\langle 001 \rangle$ and $\langle 111 \rangle$, in tension and compression; however, instead of a full 3D sample model, an RVE with isovolumetric grains and fixed boundary conditions on the RVE sides was used for the identification procedure.

\section{Results and discussion}

\subsection{Deformation under uniaxial tension}\label{Sec.4.1}
\label{dogbone}

\begin{figure}		
    \centering
    \includegraphics[angle=0,width=0.98\textwidth]{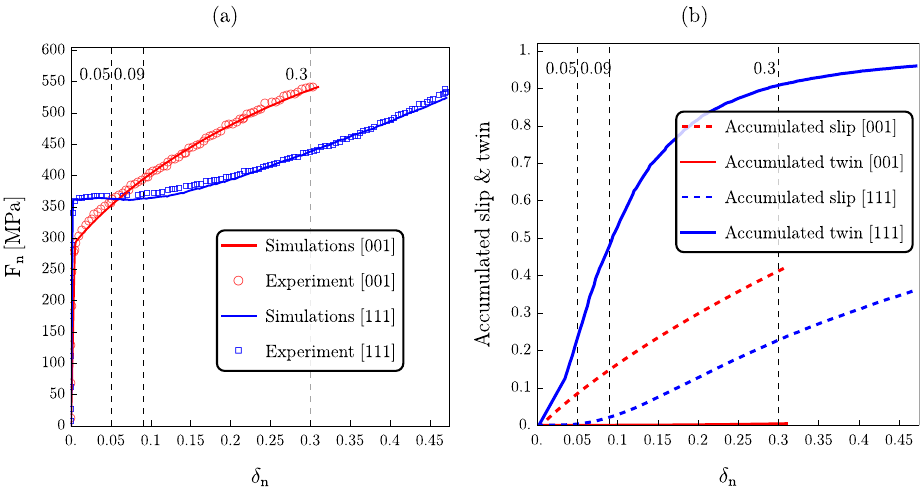}
	\caption{(a) Comparison of experimentally obtained and predicted normalized force–displacement ($F_n -\delta_n$) curves for dog-bone specimens with crystal orientations $[001]$ and $[111]$ subjected to uniaxial tension. (b) Predicted evolution of accumulated slip and twin volume fractions, averaged over the gauge region of the specimen for the two orientations.}
	\label{fig:identification}
\end{figure}

The normalized force–displacement response of dog-bone specimens with crystal orientations $[001]$ and $[111]$ subjected to uniaxial tension, as obtained from experiments and predicted using CPFE simulations, is shown in Fig.~\ref{fig:identification}(a). The normalized force is defined as $F_n = F/A_0$ and the normalized displacement as $\delta_n = \dot{\delta} t / L_0$, where $F$ is the total reaction force, $A_0$ is the initial cross-sectional area of the specimen gauge section, and $L_0$ is the initial gauge length. Although the experimental results in Fig.~\ref{fig:identification}(a) were used for the identification of constitutive parameters in Section~\ref{idpara}, the near-perfect agreement between experiments and simulations for both $[001]$- and $[111]$-oriented dog-bone specimens not only demonstrates the quality of the parameter identification, but also confirms that the constitutive model described in Section~\ref{Sec:CP} can accurately capture slip and twinning deformation mechanisms in austenitic manganese steel.

The predicted evolution of accumulated slip and twin volume fraction in the gauge section of the dog-bone specimens is shown in Fig.~\ref{fig:identification}(b). These two quantities are defined as: 
\begin{equation}\label{Eq:modact}
    {\bar{\Gamma}}= \frac{\sum_{i}\Gamma_{i} V_{i}}{\sum_{i} V_{i}}\quad
     {\bar{f}^{TW}}= \frac{\sum_{i} f^{TW}_{i} V_{i}}{\sum_{i} V_{i}}
\end{equation}
where $\Gamma_i$ is the accumulated slip, $f^{\mathrm{TW}}_i$ is the twin volume fraction, and $V_i$ is the volume associated with the $i^{\mathrm{th}}$ Gauss point. Note that if reorientation has occurred at a given Gauss point, the twin volume fraction at that point is set to one.

The results in Fig.~\ref{fig:identification}(a) clearly show that the fully calibrated CPFE simulations accurately predict the experimentally observed strain-hardening response for both $[001]$- and $[111]$-oriented specimens. Specifically, the model captures the initially high but continuously decreasing hardening rate (i.e., concave downward stress-strain curve) for the $[001]$ specimen, and the initial near-zero hardening followed by a concave upward stress–strain curve for the $[111]$ specimen. These orientation-dependent hardening behaviors reflect the underlying deformation mechanisms in the material.

As shown in Fig.~\ref{fig:identification}(b), twinning activity is negligible for the $[001]$ orientation, with deformation governed primarily by slip accumulation. In contrast, significant twinning is observed for the $[111]$ orientation. For this case, the twin volume fraction reaches approximately $0.5$ at a nominal strain of $0.09$. Beyond this point, the rate of increase in twin volume fraction slows and eventually saturates at a value of about $0.95$. At the same time, slip remains nearly inactive up to a nominal strain of $0.1$. After this point, a steady increase in accumulated slip is observed as slip becomes active within the reoriented regions of the specimen. The initial low-hardening stage in the $[111]$ stress–strain curve corresponds to twinning-dominated deformation. Once twinning saturates and slip becomes the dominant mechanism, the hardening rate begins to rise—marking the transition from initial near-zero hardening to concave upward stress–strain response for the $[111]$ orientation.

\begin{figure}
	\centering
	\includegraphics[angle=0,width=\textwidth]{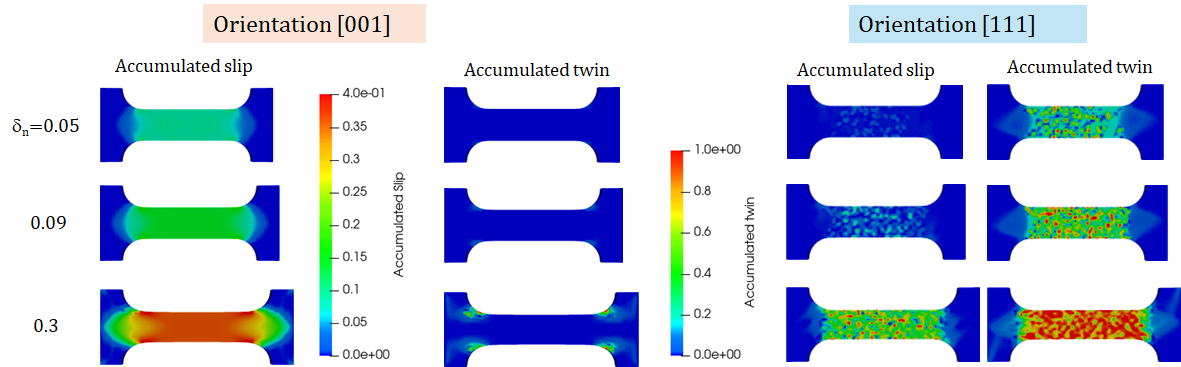}
	\caption{Contour plots showing the predicted evolution of accumulated slip and twin volume fraction in dog-bone specimens with crystal orientations $[001]$ and $[111]$ subjected to uniaxial tension. Results are shown at three levels of normalized displacement: $\delta_n = 0.05$, $0.09$, and $0.3$, which are also marked in Fig.~\ref{fig:identification}.}
	\label{fig:twinning111}
\end{figure}

The distribution of accumulated slip and twin volume fraction in the gauge section of dog-bone specimens with crystal orientations $[001]$ and $[111]$ at three levels of imposed displacement is shown in Fig.~\ref{fig:twinning111}. In the $[001]$ specimen, deformation is dominated by slip, which remains fairly uniform throughout the gauge section. However, some twinning activity is observed in the fillet region near the gauge section at larger imposed displacements. In contrast, during the early stages of deformation in the $[111]$ specimens, the gauge section is dominated by twinning, with almost no slip activity observed. At larger imposed displacements, when twinning tends to saturate, i.e., most regions in the gauge section have already twinned and reoriented, slip activity in the reoriented regions begins to dominate the deformation. Notably, due to the probabilistic, twin-volume-consistent reorientation scheme employed in this work, twinning in the gauge section occurs in discrete regions. Because the crystal plasticity formulation used here is local, it does not produce continuous twin bands spanning the width of the gauge section, as seen in the single-crystal experiments by \cite{zheng2025interaction}. Nevertheless, our results are in perfect agreement with EBSD observations reported in cite{zheng2025interaction}, which show extensive twinning activity when the material is under tension in the $[111]$ direction and hardly any such activity for tension in the $[001]$ direction.

It has also been confirmed that, in agreement with the highest Schmid factors, three twin systems---$(111)[11\bar{2}]$, $(\bar{1}11)[\bar{1}\bar{2}1]$, and $(1\bar{1}1)[\bar{2}\bar{1}1]$---are active when the tension axis is perfectly aligned with the crystallographic direction $[11\bar{1}]$, while a single twin system, $(1\bar{1}1)[\bar{2}\bar{1}1]$, is predominantly active when the tension axis is slightly misoriented with respect to the $[11\bar{1}]$ direction, as in the case of orientation $[111]$Exp. According to the reorientation scheme used in this work, the new orientation, assumed once the reorientation condition is met, corresponds to the most active twin system at the given point. For the $[111]$ case, this results in the appearance of areas with one new twin-related orientation. This finding is in agreement with the EBSD observations of \cite{zheng2025interaction}. Moreover, when observing the area fraction of twins shown in Fig. 2b of \cite{zheng2025interaction} in conjunction with the DIC maps in Fig. 3 therein, the layers of intensive strain can be correlated with the twinned regions. Using these experimental findings, it is also possible to verify that the evolution of twin volume fraction with strain shown in Fig.~\ref{fig:identification}b is correctly captured by the present approach.

\subsection{Influence of notch on deformation}
\label{notch}

\begin{figure}	
	\centering
     \includegraphics[angle=0,width=0.98\textwidth]{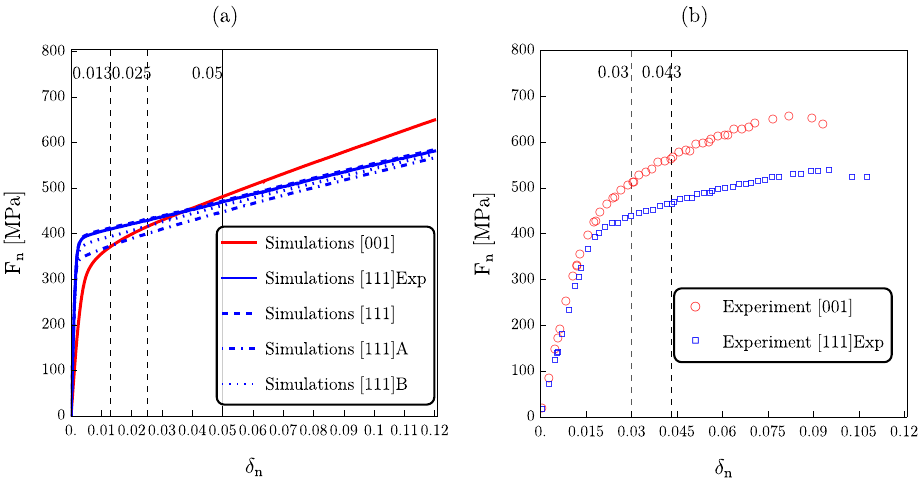}
	\caption{(a) Predicted normalized force–displacement ($F_n$–$\delta_n$) curves for single-edge notch specimens with crystal orientations $[001]$ and $[111]$. For the $[111]$ orientation, parametric studies are also conducted to assess the effect of deviations from the exact $[111]$ orientation. All orientation cases analyzed are listed in Table~\ref{tab:crystal orientations}. (b) Experimentally obtained $F_n$–$\delta_n$ curves for single-edge notch specimens with crystal orientations $[001]$ and $[111]$Exp.}
	\label{fig:notch-stress}
\end{figure}

We now analyze the effect of a notch on the orientation-dependent deformation response and mechanisms of single-crystal austenitic manganese steel. In particular, we analyze single-edge notch specimens with the loading axis aligned along either the $[001]$ or $[111]$ crystallographic directions. In the simulations, for the $[111]$ orientation, parametric studies are also conducted to assess the effect of small deviations from the exact orientation. All orientation cases analyzed are listed in Table~\ref{tab:crystal orientations}.

Figure~\ref{fig:notch-stress}(a) shows the predicted normalized force–displacement ($F_n$–$\delta_n$) curves for single-edge notch specimens, while the experimentally obtained results are shown in Fig.~\ref{fig:notch-stress}(b). As with the dog-bone specimens, $\delta_n = \dot{\delta} t / L_0$, where $L_0$ is the gauge length, and $F_n = F / A_0$, where $F$ is the total reaction force. However, for single-edge notch specimens, $A_0$ is the cross-sectional area of the ligament, i.e., the cross-sectional area of the gauge section minus the cross-sectional area removed by the notch. Note that, unlike the dog-bone specimens where extensometers were used in the experiments, the force–displacement curves for single-edge notch specimens are not corrected for machine compliance. Nevertheless, the predicted and experimental results show very good qualitative agreement. For example, both simulation and experiment show greater strain hardening in the $[001]$ specimens than in the $[111]$ specimens, and neither shows a concave-upward hardening response for the notched $[111]$ specimens. Additionally, a comparison of the $F_n$–$\delta_n$ curves for dog-bone specimens in Fig.~\ref{fig:identification}(a) and for single-edge notch specimens in Fig.~\ref{fig:notch-stress} clearly demonstrates a notch-strengthening effect in both $[001]$ and $[111]$ specimens, in both experimental and simulation results. Furthermore, simulation results from the parametric studies show that small deviations from the exact $[111]$ orientation do not significantly affect the $F_n$–$\delta_n$ response of single-edge notch specimens, apart from slight differences in the early stages of deformation.

\begin{figure}
	\centering
	\begin{tabular}{cc}
		\includegraphics[angle=0,width=0.98\textwidth]{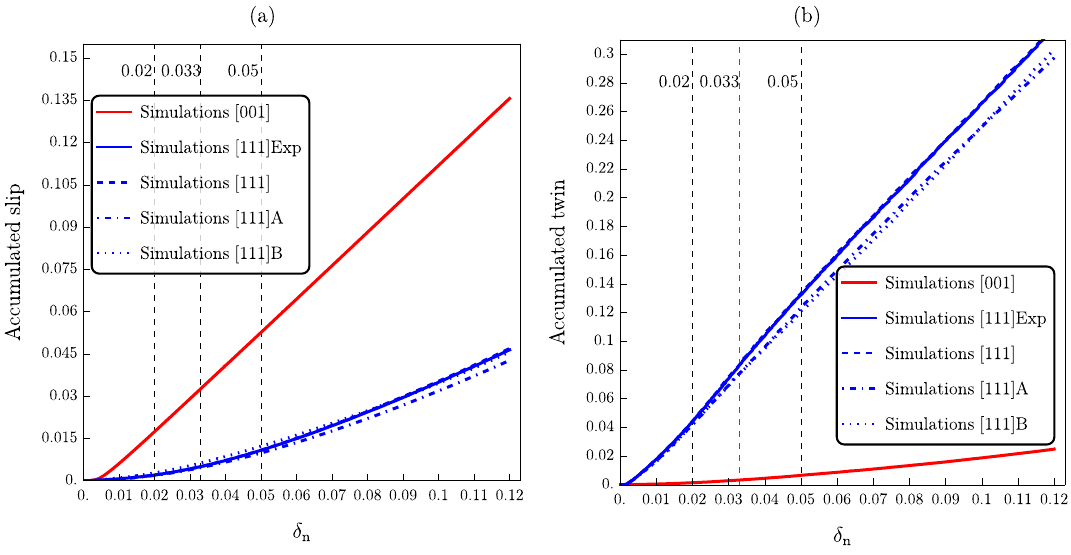}\\
	\end{tabular}
	\caption{Predicted evolution of (a) accumulated slip and (b) twin volume fraction, averaged over the gauge region of single-edge notch specimens with crystal orientations $[001]$ and $[111]$. For the $[111]$ orientation, parametric studies are conducted to assess the effect of deviations from the exact $[111]$ orientation. All orientation cases analyzed are listed in Table~\ref{tab:crystal orientations}.}
	\label{fig:notch-sliptwin}
\end{figure}

The predicted evolution of accumulated slip and twin volume fraction, averaged over the entire gauge region of the single-edge notch specimens, is shown in Fig.~\ref{fig:notch-sliptwin}(a) and (b), respectively. As in the dog-bone specimens (Fig.~\ref{fig:identification}(b)), deformation in the $[001]$ specimen is primarily governed by slip, while twinning dominates in all four $[111]$ variants. However, due to the presence of the notch, which results in a more complex stress state, a modest amount of twinning activity is also observed in the $[001]$ single-edge notch specimen. This was almost entirely absent in the $[001]$-oriented dog-bone specimen. Furthermore, in the $[111]$ single-edge notch specimens, slip activity is observed even at early stages of deformation. In contrast, in the dog-bone specimens, slip becomes active only after significant twinning and reorientation. For example, at $\delta_n = 0.05$, there is already a small but noticeable amount of accumulated slip in the $[111]$-oriented single-edge notch specimen (Fig.~\ref{fig:notch-sliptwin}(a)). At the same $\delta_n$, the accumulated slip in the $[111]$-oriented dog-bone specimen is essentially zero (Fig.~\ref{fig:identification}(b)). This increase in slip is consistent with the differences in strain-hardening behavior between the $[111]$-oriented dog-bone specimens (Fig.~\ref{fig:identification}(a)) and the single-edge notch specimens (Fig.~\ref{fig:notch-stress}(a)), particularly the absence of the early-stage non-hardening regime in the $[111]$ single-edge notch response.

\begin{figure}[h!]
	\centering
	\includegraphics[angle=0,width=\textwidth]{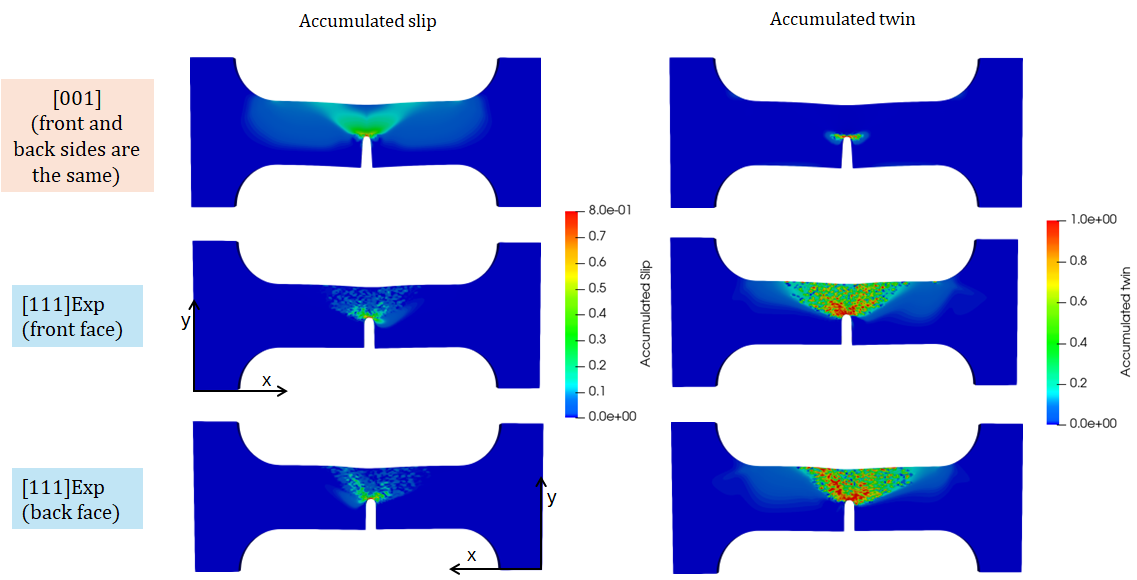}	
	\caption{Contour plots showing the predicted evolution of accumulated slip and twin volume fraction in single-edge notch specimens with crystal orientations $[001]$ and $[111]$Exp (see Table~\ref{tab:crystal orientations}) at a normalized displacement $\delta_n = 0.05$, which is also indicated in Figs.~\ref{fig:notch-stress} and \ref{fig:notch-sliptwin}.}
    \label{fig:notch-fullmaps}
\end{figure}

Contour plots of accumulated slip and twin volume fraction for the $[001]$ and $[111]$Exp specimens at $\delta_n = 0.05$ are shown in Fig.~\ref{fig:notch-fullmaps}. In both cases, slip and twinning activity are concentrated near the notch tip. Although deformation is confined to this region, twinning occurs in discrete pockets. The twin volume fraction in these pockets rapidly evolves to one. Once reoriented, these regions continue to deform by slip. This behavior explains the early onset of slip activity in the $[111]$-oriented single-edge notch specimens.

For the $[001]$-oriented specimen, the distributions of both slip and twin volume fraction are symmetric with respect to the notch. In contrast, for the $[111]$Exp-oriented specimen, both slip and twinning are clearly asymmetric across the notch. This in-plane asymmetry in twinning activity for the $[111]$ orientation is consistent with the experimental observations reported by \cite{zheng2025interaction}. As seen in Fig.~\ref{fig:notch-fullmaps}, there is also an out-of-plane asymmetry in the deformation pattern between the front and back faces of the $[111]$Exp specimen. This asymmetry arises from the inclination of the sample plane with respect to the $\{011\}$ crystal plane, which is a symmetry plane of the FCC lattice. This front–back asymmetry is not observed in the other three $[111]$ specimens analyzed computationally, where the sample plane remains aligned with $\{011\}$. However, small differences between the two faces may still occur due to the probabilistic nature of the reorientation scheme.

\subsection{Slip–twinning–notch interaction and failure implications}

\begin{figure}	
	\centering
\includegraphics[angle=0,width=0.98\textwidth]{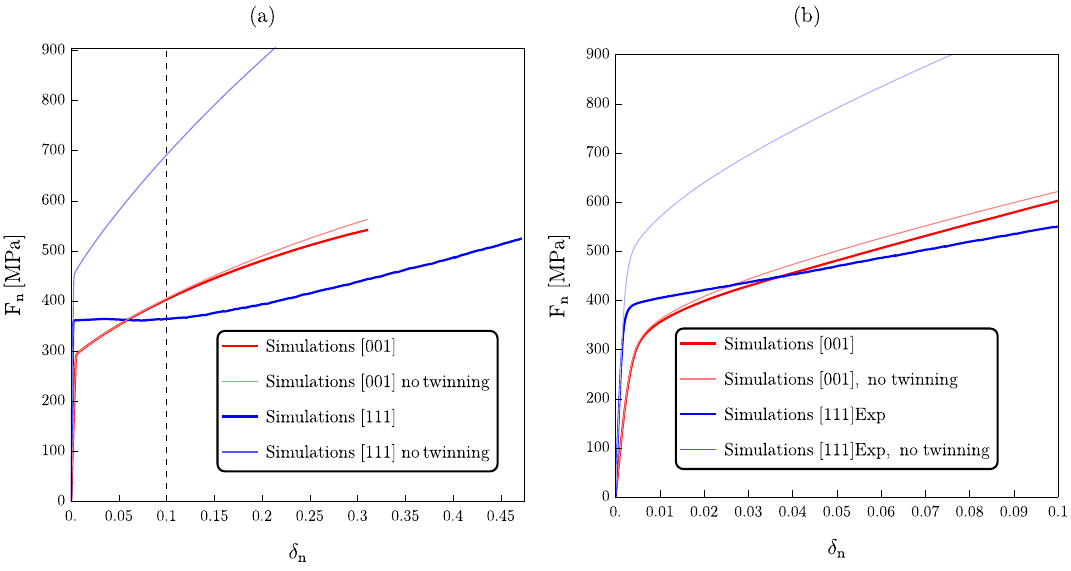}
	\caption{Predicted normalized force–displacement ($F_n$–$\delta_n$) curves for (a) dog-bone specimens with crystal orientations $[001]$ and $[111]$ and (b) single-edge notch specimens with crystal orientations $[001]$ and $[111]$Exp (see Table~\ref{tab:crystal orientations}), with and without deformation twinning.}   
	\label{fig:stress-twin-study}
\end{figure}

We now focus on understanding the interaction between the two deformation mechanisms, slip and twinning, and the effect of a notch on the orientation-dependent mechanical response of single-crystal austenitic manganese steel. To this end, we carry out simulations in which the twinning mechanism is selectively activated in both $[001]$- and $[111]$-oriented dog-bone as well as $[001]$- and $[111]$Exp-oriented single-edge notch specimens. All calculations are performed using the same set of parameters listed in Section~\ref{idpara}, and the full description of the crystallographic orientations of the specimens is given in Table~\ref{tab:crystal orientations}. For the cases without twinning, the contributions from twinning are simply disabled.

Figure~\ref{fig:stress-twin-study} presents the normalized force ($F_n$) versus normalized displacement ($\delta_n$) response of $[001]$- and $[111]$-oriented dog-bone as well as $[001]$- and $[111]$Exp-single-edge notch specimens, with and without deformation twinning included in the simulations. As seen in Fig.~\ref{fig:stress-twin-study}, the mechanical response of both $[001]$-oriented dog-bone and single-edge notch specimens is largely insensitive to whether twinning is active. A comparison of Fig.~\ref{fig:stress-twin-study}(a) and Fig.~\ref{fig:stress-twin-study}(b) also shows that the presence of a notch in the $[001]$-oriented specimen leads to a strengthening effect. This is evident from both the increase in the value of $F_n$ at the onset of nonlinearity in the $F_n$–$\delta_n$ response and the higher $F_n$ values observed during subsequent loading. This notch-strengthening effect is present at least prior to fracture, even in the experimental results, as shown in Sections~\ref{dogbone} and~\ref{notch}.

In contrast, as seen in Fig.~\ref{fig:stress-twin-study}, activation of twinning has a pronounced effect on the $F_n$–$\delta_n$ response of both $[111]$-oriented dog-bone and $[111]$Exp-oriented single-edge notch specimens. In both cases, when twinning is disabled, the value of $F_n$ at the onset of nonlinearity is significantly higher, and higher $F_n$ values are observed throughout subsequent loading. Importantly, in the absence of twinning, the initially non-hardening (nearly flat $F_n$–$\delta_n$) response is entirely absent in the $[111]$Exp-oriented dog-bone specimen. Additionally, without twinning, the $[111]$-oriented dog-bone and $[111]$Exp-oriented single-edge notch specimens exhibit the onset of nonlinearity at an $F_n$ value approximately 1.5 times greater than that of the corresponding $[001]$-oriented specimens. This behavior is consistent with the relative Schmid factors associated with these orientations. However, when twinning is active, the difference in the $F_n$ value at the onset of nonlinearity between the $[111]$- and $[001]$-oriented specimens is relatively small.

\begin{figure}
	\centering
	\includegraphics[angle=0,width=0.8\textwidth]{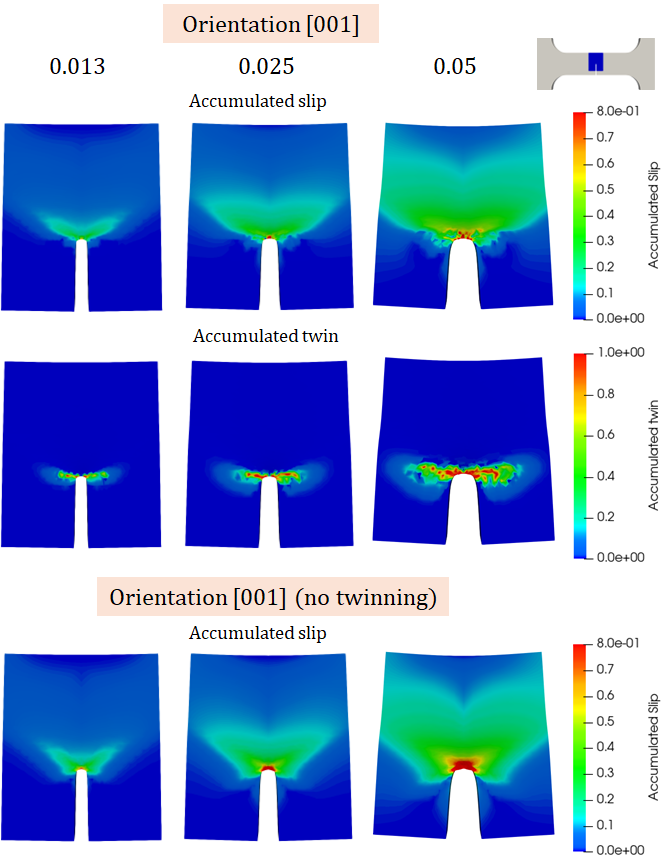}
	\caption{Contour plots showing the predicted evolution of accumulated slip and twin volume fraction in $[001]$-oriented single-edge notch specimen at three normalized displacements $\delta_n = 0.013, 0.025,$ and $0.05$. Also shown are the contour plots of accumulated slip in $[001]$-oriented single-edge notch specimen at the same three normalized displacements for a scenario where twinning is assumed to be absent.}
    \label{fig:contours-evol-001}
\end{figure}

\begin{figure}
	\centering
	\includegraphics[angle=0,width=0.8\textwidth]{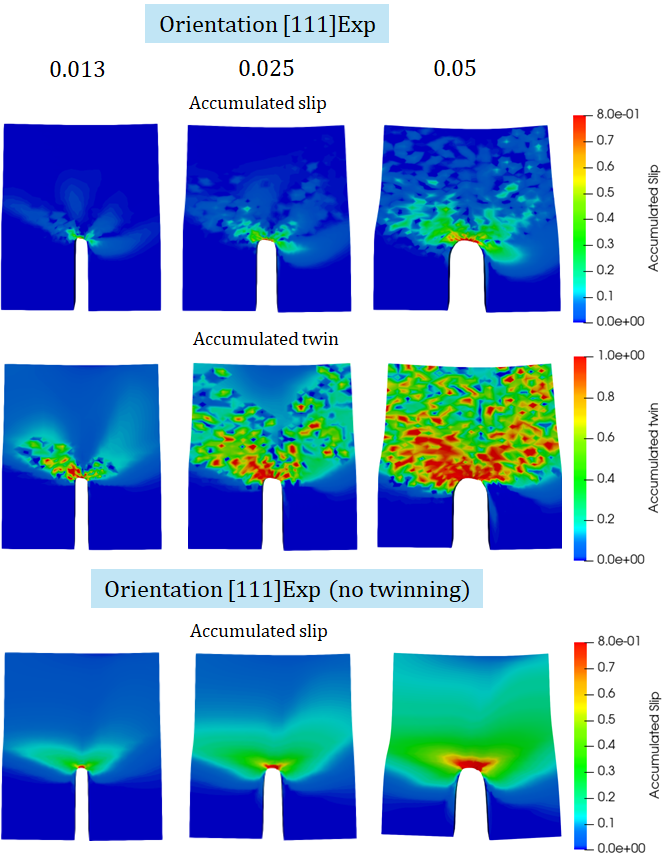}
	\caption{Contour plots showing the predicted evolution of accumulated slip and twin volume fraction in $[111]$Exp-oriented single-edge notch specimen at three normalized displacements $\delta_n = 0.013, 0.025,$ and $0.05$. Also shown are the contour plots of accumulated slip in $[111]$Exp-oriented single-edge notch specimen at the same three normalized displacements for a scenario where twinning is assumed to be absent.}
    \label{fig:contours-evol-111}
\end{figure}

The distribution of accumulated slip and twin volume fraction in the $[001]$- and $[111]$Exp-oriented single-edge notch specimens at the same three imposed displacements is shown in Figs.~\ref{fig:contours-evol-001} and \ref{fig:contours-evol-111}, respectively. In both figures, results are also shown for scenarios where twinning is assumed to be absent. In both the $[001]$- and $[111]$Exp-oriented single-edge notch specimens, deformation is concentrated near the notch. In the $[001]$-oriented specimen, deformation is symmetric across the notch and is largely governed by accumulated slip. A narrow and relatively small zone surrounding the notch in the $[001]$-oriented specimen does exhibit twinning, which arises due to the multiaxial stress state that develops near the notch tip. However, the overall distribution of accumulated slip in the $[001]$-oriented single-edge notch specimen is similar with and without twinning. The only difference is that, since twinning in the simulations occurs in discrete regions, it introduces some additional spatial discreteness in the slip distribution near the notch tip.  

In the $[111]$Exp-oriented single-edge notch specimen (Fig.~\ref{fig:contours-evol-111}), deformation is largely governed by twinning, and the distribution of both accumulated slip and twin volume fraction is asymmetric across the notch. The accumulated slip is concentrated in three distinct regions. Two of these correspond to areas where twinning is also dominant. However, one region—located behind the notch tip and below the notch (on the right side of the notch in Fig.~\ref{fig:contours-evol-111})—exhibits significant slip activity without any observed twinning. This slip-dominated region contributes to the flattening of the initially circular notch tip. While the distribution of accumulated slip remains asymmetric even in the absence of twinning, the additional spatial discreteness associated with twinning reorientation is absent, and the deformation of the notch tip is comparatively less pronounced. Although not shown here, the other $[111]$ orientations (see Table~\ref{tab:crystal orientations}) also qualitatively show similar distributions of accumulated slip and twinning near the deforming notch.

\begin{figure}[h!]
	\centering	
	\small{(a)}\\
	\includegraphics[angle=0,width=0.4\textwidth]{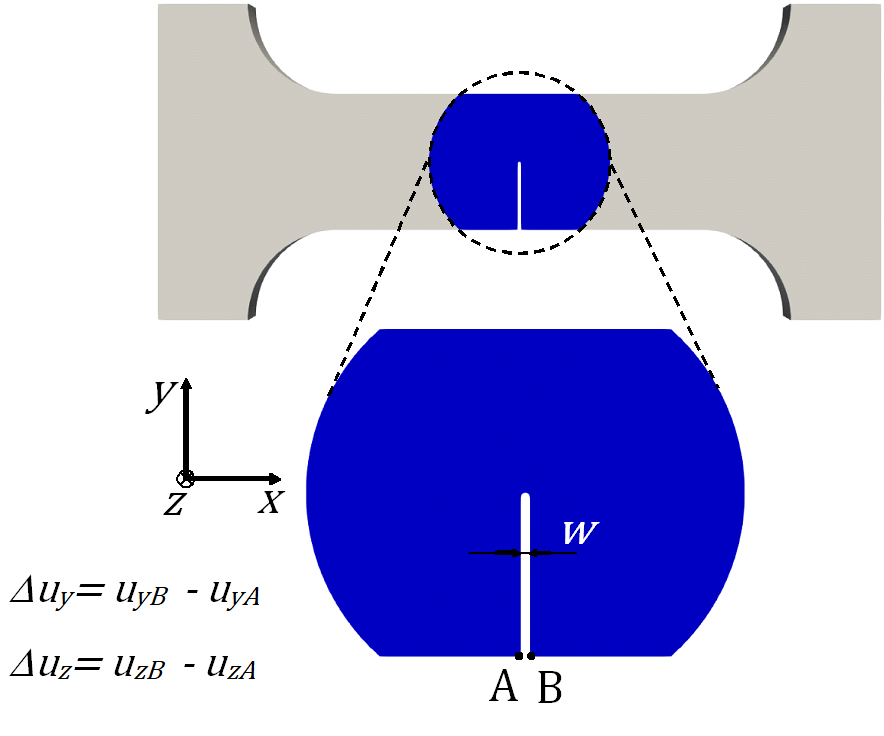}\\		
    \includegraphics[angle=0,width=0.98\textwidth]{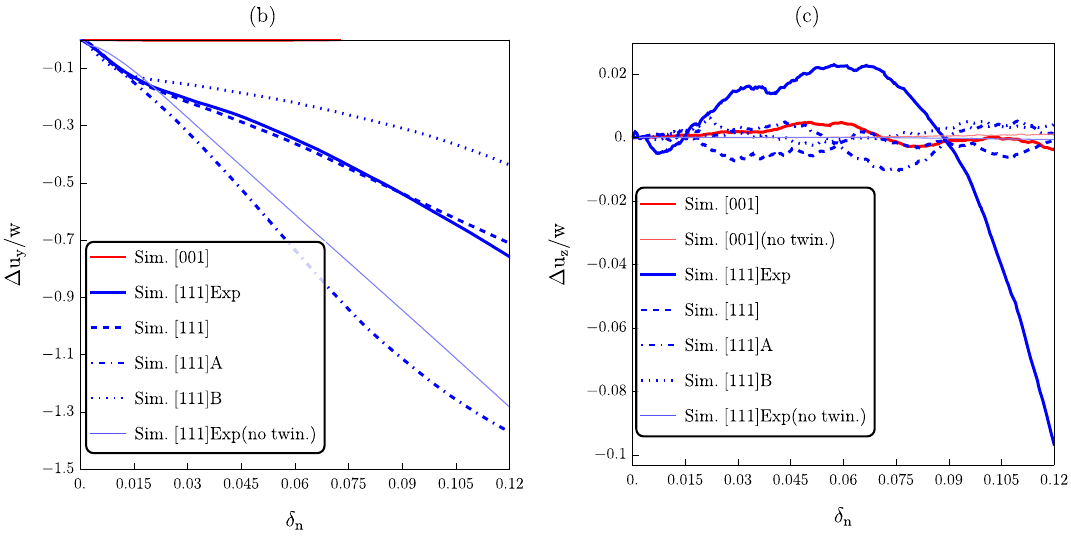}
	\caption{(a) Schematic illustrating two parameters, $\Delta u_y$ and $\Delta u_z$, used to characterize asymmetric notch opening in single-edge notch specimens. Evolution of predicted (b) normalized $\Delta u_y$ and (c) normalized $\Delta u_z$ with normalized displacement $\delta_n$ for various orientations of the single-edge notch specimens. All orientation cases analyzed are listed in Table~\ref{tab:crystal orientations}. Results are also shown for the $[111]$Exp-oriented specimen under the assumption that twinning is absent.}
    \label{fig:notch-asymmetry}
\end{figure}

The asymmetric deformation field in the $[111]$Exp-oriented notched specimen, as shown in Fig.~\ref{fig:contours-evol-111}, leads to asymmetric notch deformation. This notch asymmetry is further analyzed in Fig.~\ref{fig:notch-asymmetry}. The plots in Figs.~\ref{fig:notch-asymmetry}(b) and \ref{fig:notch-asymmetry}(c) show the differences in the displacement components $u_y$ and $u_z$, respectively, between two nodes A and B located on either side of the notch, as schematically illustrated in Fig.~\ref{fig:notch-asymmetry}(a). These displacement differences are normalized by the initial notch width, $w = 90\,\mu$m. In the case of a perfectly symmetric notch opening, the normalized values would be zero.

The in-plane asymmetry of the notch opening, quantified by $\Delta u_y / w$ in Fig.~\ref{fig:notch-asymmetry}(b), varies among the different orientations. It is negligible for the $[001]$-oriented specimen and is strongly influenced by the misorientation of the $[111]$ direction with respect to the specimen axis in the four other $[111]$ cases. The largest asymmetry is observed for the $[111]$A specimen, in which the $[111]$ direction is rotated by $6^\circ$ clockwise about the $z$-axis. A similar level of asymmetry is also observed when twinning is disabled, indicating that this effect primarily arises from crystallographic anisotropy.

As previously discussed in relation to Fig.~\ref{fig:notch-fullmaps}, the asymmetry in the $z$-direction, shown in Fig.~\ref{fig:notch-asymmetry}(c), is much smaller than the asymmetry in the $y$-direction. The $z$-direction asymmetry is observed primarily for the $[111]$Exp orientation, in which the sample plane is misaligned with respect to the $(\bar{1}10)$ crystallographic plane. Even in this case, however, the ratio $\Delta u_z / w$ is at least an order of magnitude smaller than $\Delta u_y / w$. For the other orientations, fluctuations in $\Delta u_z / w$ are nearly two orders of magnitude smaller and are attributed to the probabilistic nature of the twin reorientation scheme. As expected, when twinning is disabled, the value of $\Delta u_z / w$ is exactly zero.

\begin{figure}
	\centering
	\includegraphics[angle=0,width=0.9\textwidth]{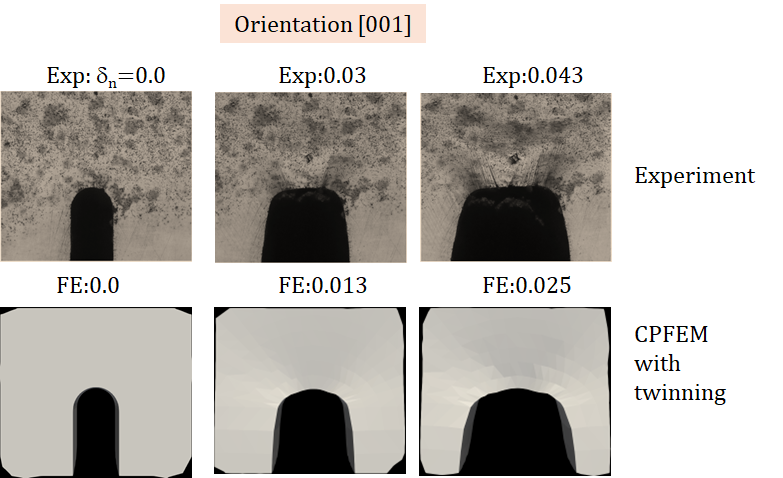}\\    
	\includegraphics[angle=0,width=0.9\textwidth]{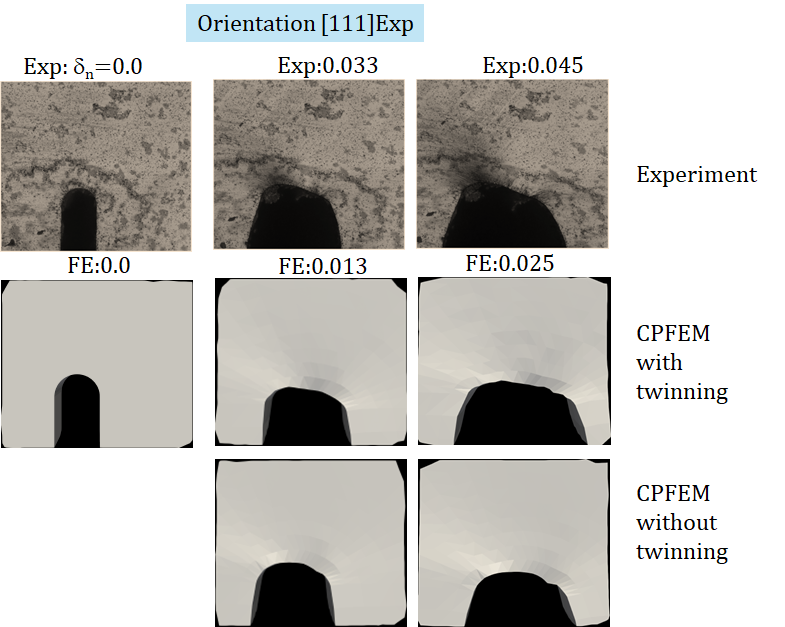}
	\caption{Comparison of experimentally observed and predicted notch root shapes in $[001]$ and $[111]$Exp-oriented single-edge notch specimens (see Table~\ref{tab:crystal orientations}) at two values of the imposed normalized displacement ($\delta_n$). For the $[111]$Exp-oriented specimen, predictions are also shown for a scenario where twinning is assumed to be absent. The region shown in both experimental images and predictions spans approximately $330\ \mu$m (height) $\times$ $375\ \mu$m (width).}
    \label{fig:notch-def}
\end{figure}

In Fig.~\ref{fig:notch-def}, we compare the deformed shape of the notch in both $[001]$- and $[111]$Exp-oriented single-edge notch specimens as observed in experiments and as predicted by simulations. The results are also shown for the $[111]$Exp-oriented specimen under a scenario where twinning is assumed to be absent. For a meaningful comparison, all results are presented at approximately the same current notch width, since the experimentally measured normalized displacement ($\delta_n$) of the entire specimen is not directly comparable to simulation values due to machine compliance effects in the experimental setup. The values of $\delta_n$ indicated in the figure for both the experiments and the simulations correspond to those marked in Figs.~\ref{fig:notch-stress}(a) and (b), respectively.

As seen in the figure, there is very good agreement between the experimentally observed and simulated notch shapes. For the $[001]$-oriented specimen, consistent with experimental observations, the simulations predict that the initially semi-circular notch tip flattens and opens symmetrically. For the $[111]$Exp-oriented specimen, the notch tip also flattens, but the flattened edge is inclined toward one side, resulting in a highly asymmetric overall notch deformation. Predictions for the $[111]$Exp-oriented specimen without twinning show that the deformation of the notch remains asymmetric, although the extent of asymmetry and notch tip flattening is significantly reduced compared to the case with active twinning.

The asymmetric deformation of an initially symmetric discontinuity in FCC crystals has also been observed in other contexts. For example, in the work by \cite{SRIVASTAVA201510}, which focused on the effect of crystal orientation on the evolution of initially spherical voids in FCC single crystals subjected to tensile creep loading, it was found that when the maximum principal stress direction was aligned with the $⟨111⟩$ crystallographic direction, both the matrix and the initially spherical void deformed asymmetrically, even under high imposed stress triaxiality (defined as the ratio of the first to the second stress invariants). The void evolved into a polygonal shape with rounded corners, highlighting the inherent anisotropy-driven asymmetry in crystal plasticity.

\begin{figure}[h!]
	\centering
	\includegraphics[angle=0,width=0.95\textwidth]{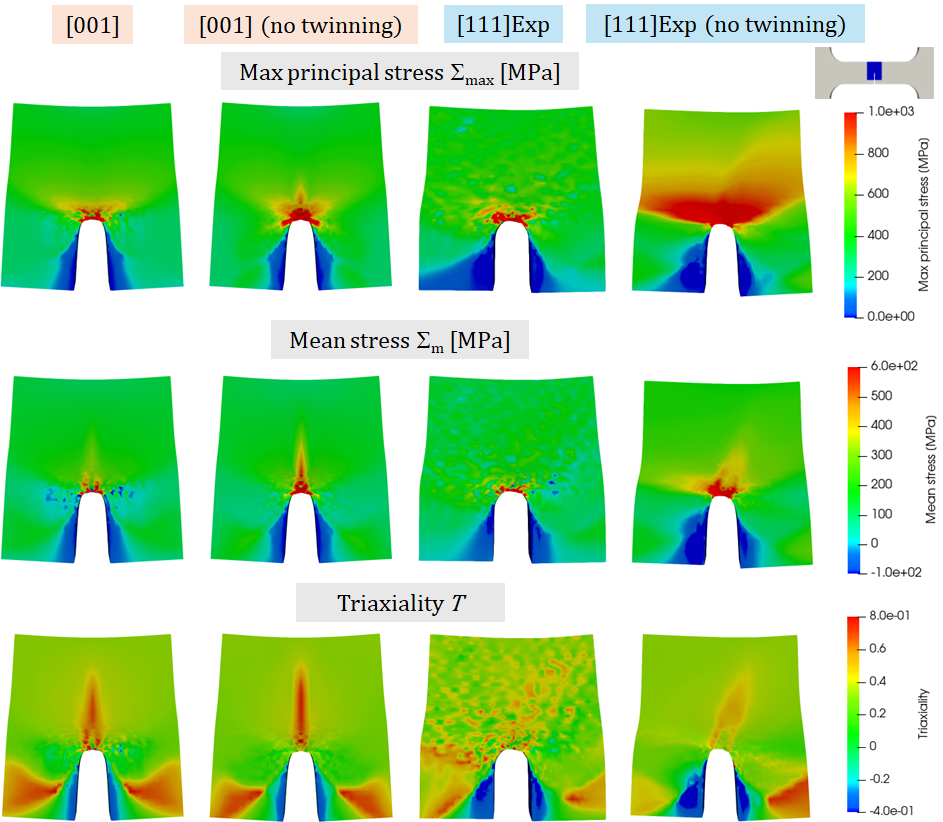}
	\caption{Contour plots showing the predicted distribution of maximum principal stress ($\Sigma_{\mathrm{max}}$), mean stress ($\Sigma_{m}$), and triaxiality ($T$) at a normalized displacement $\delta_n = 0.05$ in single-edge notch specimens with crystal orientations $[001]$ and $[111]$Exp (see Table~\ref{tab:crystal orientations}), with and without deformation twinning.}
    \label{fig:stress-05}
\end{figure} 

In the experiments of \cite{zheng2025interaction}, it was observed that for $[001]$-oriented single-edge notch specimens, a single crack nucleated from the center of the symmetrically deforming notch, whereas for $[111]$Exp-oriented specimens, multiple microcracks nucleated along the periphery of the asymmetrically deforming notch. Although crack nucleation is not explicitly modeled in this work, Fig.~\ref{fig:stress-05} presents the predicted distribution of maximum principal stress ($\Sigma_{\mathrm{max}}$), mean stress ($\Sigma_{m}$), and stress triaxiality ($T$) in both $[001]$- and $[111]$Exp-oriented single-edge notch specimens under scenarios where deformation twinning is either active or suppressed. These parameters provide insight into the propensity for crack initiation.

As shown in the figure, for the $[001]$-oriented specimen, with or without twinning, the distributions of $\Sigma_{\mathrm{max}}$, $\Sigma_{m}$, and $T$ are nearly symmetric across the notch, and their peak values—particularly those of $T$—are concentrated along the centerline of the symmetrically deforming notch. This suggests a tendency for the nucleation of a single ductile crack from the notch center. In contrast, for the $[111]$Exp-oriented specimen, again with or without twinning, the distributions of $\Sigma_{\mathrm{max}}$, $\Sigma_{m}$, and $T$ are clearly asymmetric across the notch. Their maxima, in the case with twinning, are not concentrated in a single region along the centerline, as observed in the $[001]$ case. Of particular importance is the distribution of $T$ around the deformed notch in the $[111]$Exp specimen, which shows elevated values in discrete pockets, including along the periphery of the asymmetrically deformed notch. This spatial pattern suggests a greater propensity for the nucleation of multiple ductile cracks along the notch periphery in the $[111]$Exp-oriented case.

\section{Summary and concluding remarks}

This work is motivated by the recent in-situ experimental observations by \cite{zheng2025interaction}, in which the deformation and fracture behavior of an austenitic manganese steel was examined using $\langle 001 \rangle$- and $\langle 111 \rangle$-oriented single-crystal dog-bone and single-edge notch tension specimens. To explain these observations and deconvolute the interaction of slip, twinning, and notch, we carried out detailed finite element simulations using a rate-dependent crystal plasticity framework that incorporates both crystallographic slip and deformation twinning. Limited experiments on single-crystal dog-bone and single-edge notch specimens, with loading axes aligned close to $\langle 001 \rangle$ or $\langle 111 \rangle$, were also conducted to aid the simulations. Several features of the experimental observations are accurately captured in the simulations, and the underlying mechanisms are clarified. The key findings of this study are summarized below:

\begin{itemize}

\item The fully calibrated crystal plasticity finite element (CPFE) simulations accurately reproduce the distinct stress–strain responses observed experimentally for $\langle 001 \rangle$- and $\langle 111 \rangle$-oriented dog-bone specimens. In particular, the simulations quantitatively capture the initially high strain-hardening followed by a concave downward response in the $\langle 001 \rangle$ specimens, and the initial near-zero strain-hardening followed by a concave upward response in the $\langle 111 \rangle$ specimens.

\item Consistent with experiments, the simulations show negligible twinning activity in $\langle 001 \rangle$-oriented dog-bone specimens and twinning-dominated deformation in the $\langle 111 \rangle$-oriented ones. The accumulated slip increases monotonically in the $\langle 001 \rangle$ specimens, while in the $\langle 111 \rangle$ specimens, slip becomes active only after twinning saturates. Thus, the initial near-zero hardening stage corresponds to the twinning-dominated regime, and since the twinned (and reoriented) regions require higher stresses to slip, the response transitions into a concave upward stress–strain curve.

\item For both dog-bone and single-edge notch specimens, the $\langle 001 \rangle$ orientation is initially softer than $\langle 111 \rangle$. However, due to a higher initial strain-hardening rate, the $\langle 001 \rangle$ specimens become harder with increasing deformation. In $\langle 111 \rangle$ dog-bone specimens, since the stress–strain curve transitions into a concave upward shape, they ultimately exhibit strength levels comparable to or greater than those of the $\langle 001 \rangle$ specimens.

\item  Compared to dog-bone specimens, both $\langle 001 \rangle$ and $\langle 111 \rangle$ notched specimens show a notch-strengthening effect, with higher stresses required at the same deformation levels. The stress–strain curves of both $\langle 001 \rangle$ dog-bone and single-edge notch specimens are qualitatively similar, exhibiting a nearly concave downward shape. However, the stress–strain curves of $\langle 111 \rangle$ dog-bone and single-edge notch specimens differ significantly. In the presence of a notch, the $\langle 111 \rangle$ orientation does not exhibit the near-zero hardening stage but instead shows a continuous sub-linear hardening response.

\item Comparison of predictions with and without twinning as an active deformation mechanism shows that the response of $\langle 001 \rangle$-oriented specimens remains largely unchanged, even though limited twinning may occur in single-edge notch specimens due to the multiaxial stress state. In contrast, deactivating twinning has a significant effect on the response of $\langle 111 \rangle$ specimens. Without twinning, both $\langle 111 \rangle$-oriented dog-bone and single-edge notch specimens are much harder than the $\langle 001 \rangle$ specimens, consistent with the relative Schmid factors associated with these orientations. Additionally, without twinning, both $\langle 111 \rangle$-oriented dog-bone and single-edge notch specimens exhibit a sub-linear hardening response.        

\item The CPFE simulations also accurately capture the experimentally observed notch deformation in both $\langle 001 \rangle$- and $\langle 111 \rangle$-oriented single-edge notch specimens. For the $\langle 001 \rangle$-oriented specimens, the initially semi-circular notch tip flattens and opens symmetrically. In the $\langle 111 \rangle$-oriented specimens, the notch tip also flattens, but the flattened edge becomes inclined, resulting in a highly asymmetric overall notch deformation. Predictions for the $\langle 111 \rangle$-oriented specimens without twinning show that the notch deformation remains asymmetric, although both the extent of notch tip deformation and the degree of deformation asymmetry are significantly reduced compared to the case with active twinning.

\item Although crack nucleation is not explicitly modeled, the predicted distributions of stress measures associated with ductile failure provide insight into crack initiation. For the $\langle 001 \rangle$-oriented specimens, with or without twinning, the stress fields are nearly symmetric across the notch, with peak values, especially stress triaxiality, concentrated along the centerline. This suggests the nucleation of a single crack, consistent with experiments. In contrast, the $\langle 111 \rangle$-oriented specimens show asymmetric stress distributions, and with twinning, elevated stress triaxiality appears in discrete pockets along the notch periphery. This pattern indicates a propensity for multiple crack nucleation sites, in agreement with experimental observations.

\end{itemize}

\section*{CRediT authorship contribution statement}
\textbf{S. Virupakshi:} Data curation, Formal analysis, Investigation, Methodology, Software, Validation, Visualization, Writing – original draft. \textbf{X. Zheng:} Investigation, Methodology, Validation, Visualization, Writing – review and editing. \textbf{K. Frydrych:} Methodology, Software, Writing – review and editing. \textbf{I. Karaman:} Resources, Validation, Writing – review and editing. \textbf{A. Srivastava:} Conceptualization, Data curation, Funding acquisition, Methodology, Resources,  Supervision, Validation, Writing – original draft. \textbf{K. Kowalczyk-Gajewska:} Conceptualization, Data curation, Formal analysis, Funding acquisition, Investigation, Methodology, Resources, Software, Supervision, Validation, Writing – original draft.

\section*{Declaration of competing interest}
The authors declare that they have no known competing financial interests or personal relationships that could have appeared to influence the work reported in this paper.

\section*{Acknowledgements}
A.S. acknowledges the financial support provided by the U.S. National Science Foundation grant CMMI-1944496. S.V. and K.K.-G. have been partially supported by the project No. 2021/41/B/ST8/03345 of the National Science Centre, Poland. K.K.-G. also acknowledges the financial support provided by the European Union's  Research and Innovation Program Horizon Europe under the Grant Agreement No. 101086342 – HORIZON-MSCA-2021-SE-01 Project DIAGONAL (Ductility and fracture toughness analysis of Functionally Graded Materials). The authors acknowledge that the characterization part of this work was performed in Texas A\&M University Materials Characterization Core Facility (RRID:SCR\_022202). 

\appendix

\section{Mesh sensitivity study for the PTVC reorientation scheme\label{app1}}

This appendix examines the sensitivity of the PTVC reorientation scheme to mesh size and small perturbations of parameters.

In multiple calculations related to parameter identification using a genetic algorithm, it was observed that the results of the PTVC reorientation procedure are not highly sensitive (in the ergodic sense) to small perturbations of material parameters. For a single Gauss point, even a small variation of material parameters can drastically change the moment during the calculation when the reorientation condition is satisfied, due to the random polling included in the procedure. However, when observing a statistically meaningful number of Gauss points from a given domain, their collective response is not visibly different. Despite the limitation regarding the prediction of morphological features, this is consistent with physical reality, in which the exact location of twin nucleation sites in a single crystal under uniform loading cannot be precisely predicted and is dictated by small-scale fluctuations in the crystal lattice \citep{Huang25}.

\begin{figure}
	\centering
	\includegraphics[angle=0,width=0.85\textwidth]{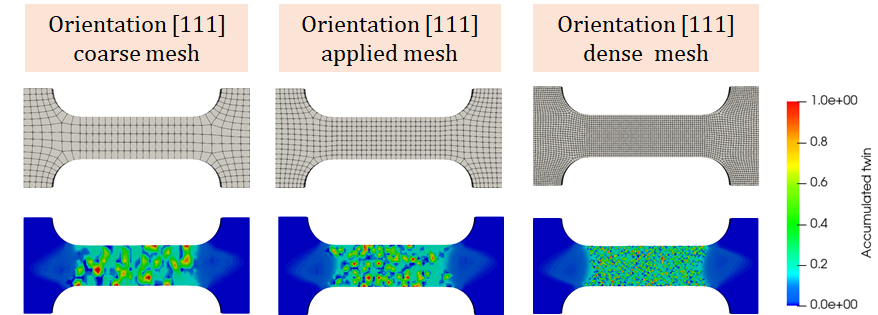}
	\caption{Three mesh densities for the dog-bone sample, together with contour plots of accumulated twin at a strain of 0.051, demonstrating the impact of the probabilistic reorientation condition, in which reorientation is performed at the Gauss point level (red spots in the contour maps correspond to reoriented domains).}
	\label{fig:R2}
\end{figure}

Mesh size influences the size of a single reoriented domain, visible as a red spot in Figs.~\ref{fig:twinning111}, \ref{fig:notch-fullmaps}, \ref{fig:contours-evol-001}, and \ref{fig:contours-evol-111}, which correspond to twinned domains. The reorientation condition is checked at each Gauss point, which then assumes the twin-related orientation. This effect is illustrated in Fig.~\ref{fig:R2}, where three different meshes---coarse, regular (used in the study in Section~\ref{Sec.4.1}), and dense---are used in the simulations. The overall effect, however, is negligible as long as a statistically meaningful number of elements is used.

In Fig.~\ref{fig:R3}, the stress–strain curves and accumulated slip and twin volume fractions predicted for the gauge area of the sample are compared for three mesh densities. Only slight differences exist between the results for the coarse and regular meshes, while the curves for the regular and dense meshes almost coincide. In Fig.~\ref{fig:R4}, the influence of mesh size on the predicted notch deformation and the spatial distribution of accumulated twinning is presented. Similar to the dog-bone sample, the notch deformation is not significantly affected when changing from the currently applied mesh to a coarser one. The normalized force–normalized displacement curves and accumulated slip and twin (twin kinetics) predicted for the gauge area of the notch sample (results not shown) show almost no difference between the coarse and regular meshes.

\begin{figure}
	\centering
	\includegraphics[angle=0,width=0.9\textwidth]{Hadfield_dogbone_mesh.pdf}
	\caption{(a) Comparison of experimentally obtained and predicted normalized force–displacement ($F_n$–$\delta_n$) curves for dog-bone specimens with the $[111]$ crystal orientation, subjected to uniaxial tension, using three different mesh densities. (b) Predicted evolution of accumulated slip and twin volume fractions, averaged over the gauge region of the specimen.}
	\label{fig:R3}
\end{figure}

\begin{figure}
	\centering
	\includegraphics[angle=0,width=\textwidth]{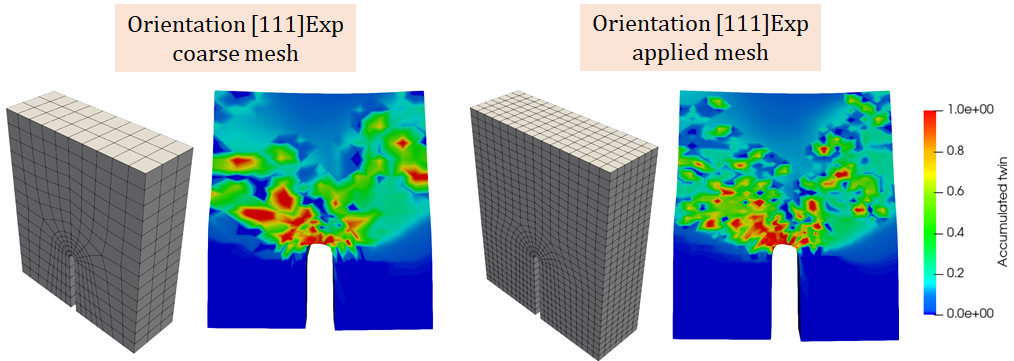}
	\caption{Two mesh densities for the notch sample, together with contour plots of accumulated twin at a strain of 0.025, demonstrating the impact of the probabilistic reorientation condition, in which reorientation is performed at the Gauss point level (red spots in the contour maps correspond to reoriented domains).}
	\label{fig:R4}
\end{figure}

\section{Influence of latent hardening coefficients\label{app2}}

Within the identification procedure, the latent hardening coefficients $q^{(\alpha\beta)}$ were assumed to be equal to 1, which was also beneficial from a computational efficiency standpoint. Naturally, the pre-assumed values of these parameters can influence the final results. To assess this effect, calculations were performed for a dog-bone specimen without a notch, oriented along $[111]$, using three alternative values of the latent hardening coefficient: $q = 0$, $q = 1.4$, and a case with latent hardening completely neglected (note that setting $q^{(\alpha\beta)} = 0$ in the formula $h_{rq}^{(\alpha\beta)} = q^{(\alpha\beta)} + (1 - q^{(\alpha\beta)})|\mathbf{n}^r \cdot \mathbf{n}^q|$ does not fully eliminate latent hardening). All other parameters were kept the same.

\begin{figure}
	\centering
	\includegraphics[angle=0,width=0.9\textwidth]{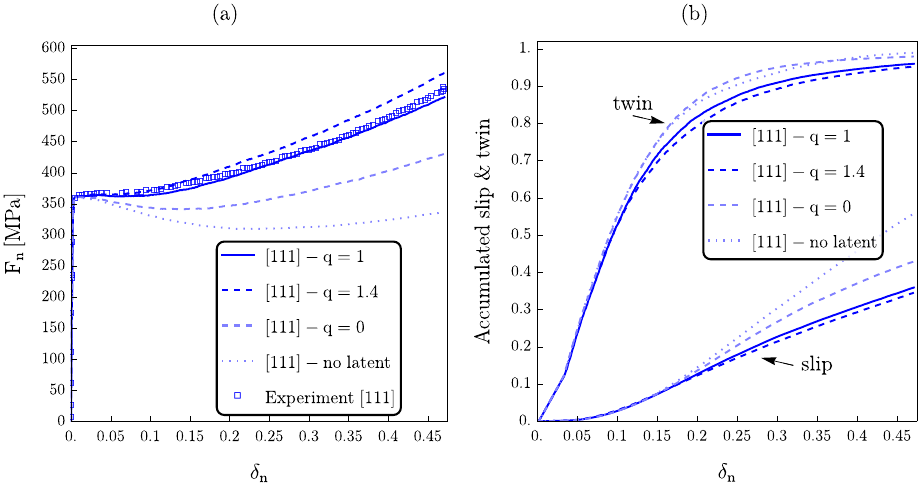}
	\caption{Comparison of results obtained with different values of $q = q^{(ss)} = q^{(ts)} = q^{(tt)}$ in Eqs.~(9) and (12): (a) normalized force–displacement ($F_n$–$\delta_n$) curves for a dog-bone specimen with the $[111]$ crystal orientation subjected to uniaxial tension; (b) predicted evolution of accumulated slip and twin volume fractions, averaged over the gauge region of the specimen.}
	\label{fig:R7}
\end{figure}

As expected, taking the latent hardening coefficient equal to zero reduced the hardening rate in the stress–strain curve (see Fig.~\ref{fig:R7}a). The impact on twin kinetics was less pronounced (see Fig.~\ref{fig:R7}b), although twins accumulated faster when $q = 0$ or when latent hardening was entirely neglected. Regarding slip activity, reducing the latent hardening coefficient increased slip activity at later stages of deformation, when slip primarily occurred in reoriented grains. In contrast, assuming a more physically justified value of $q$ greater than 1 (here $q = 1.4$) did not significantly change the results.

\begin{figure}
	\centering
	\includegraphics[angle=0,width=0.9\textwidth]{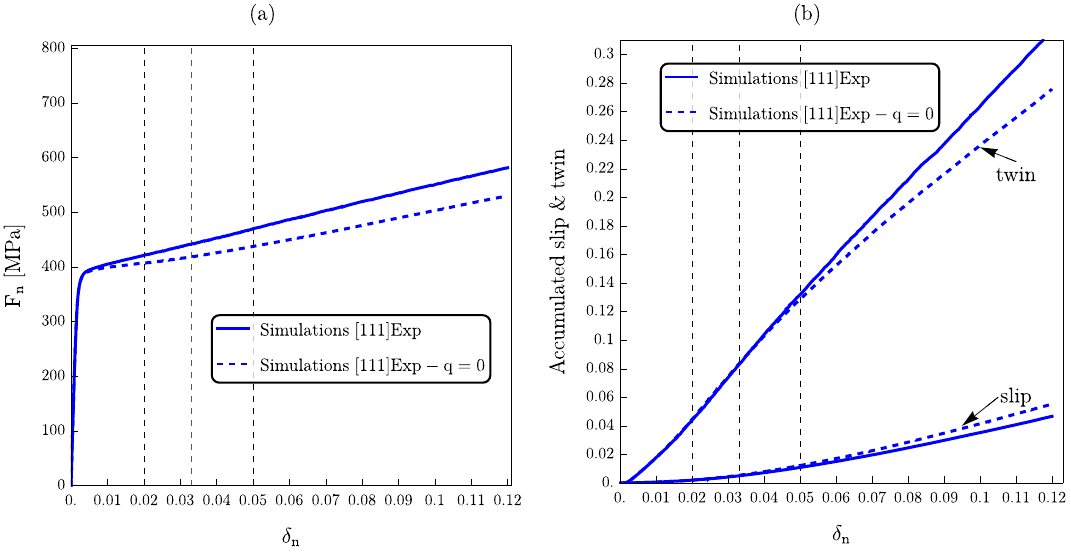}
	\caption{Comparison of results obtained with $q = 1$ and $q = 0$ for a notch sample: (a) normalized force–displacement ($F_n$–$\delta_n$) curves for notch specimens with the $[111]$Exp crystal orientation subjected to uniaxial tension; (b) predicted evolution of accumulated slip and twin volume fractions, averaged over the gauge region of the notch specimen.}
	\label{fig:R8}
\end{figure}

\begin{figure}
	\centering
	\includegraphics[angle=0,width=0.8\textwidth]{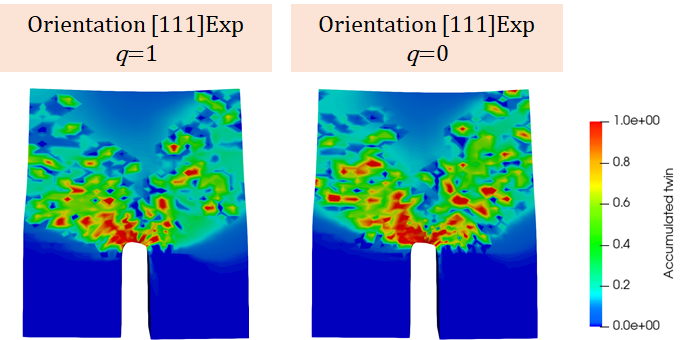}
	\caption{Comparison of contour plots of accumulated twin in the notch area at a strain of 0.025 for $q = 1$ and $q = 0$.}
	\label{fig:R9}
\end{figure}

As expected, modifying the latent hardening coefficients $q^{(\alpha\beta)}$ led to a loss of agreement between the simulated stress–strain curves and experimental data, since the parameters were originally identified for $q = 1$. Figs.~\ref{fig:R8} and \ref{fig:R9} compare results for a notch specimen with $q = 1$ and $q = 0$. As expected, the response for $q = 0$ is softer than for $q = 1$. In this case, twin kinetics are affected at later stages of the tensile process, and—unlike in the dog-bone specimen without a notch—less twinning is accumulated for $q = 0$.

\bibliography{myreferX}

\begin{thebibliography}{76}
\expandafter\ifx\csname natexlab\endcsname\relax\def\natexlab#1{#1}\fi
\providecommand{\url}[1]{\texttt{#1}}
\providecommand{\href}[2]{#2}
\providecommand{\path}[1]{#1}
\providecommand{\DOIprefix}{doi:}
\providecommand{\ArXivprefix}{arXiv:}
\providecommand{\URLprefix}{URL: }
\providecommand{\Pubmedprefix}{pmid:}
\providecommand{\doi}[1]{\href{http://dx.doi.org/#1}{\path{#1}}}
\providecommand{\Pubmed}[1]{\href{pmid:#1}{\path{#1}}}
\providecommand{\bibinfo}[2]{#2}
\ifx\xfnm\relax \def\xfnm[#1]{\unskip,\space#1}\fi
\bibitem[{Adler et~al.(1986)Adler, Olson \& Owen}]{adler1986strain}
\bibinfo{author}{Adler, P.}, \bibinfo{author}{Olson, G.}, \&
  \bibinfo{author}{Owen, W.} (\bibinfo{year}{1986}).
\newblock \bibinfo{title}{Strain hardening of hadfield manganese steel}.
\newblock {\it \bibinfo{journal}{Metall. Mater. Trans.}\/},  {\it
  \bibinfo{volume}{17}\/}, \bibinfo{pages}{1725--1737}.
\bibitem[{Ardeljan et~al.(2015)Ardeljan, McCabe, Beyerlein \&
  Knezevic}]{Ardeljan15}
\bibinfo{author}{Ardeljan, M.}, \bibinfo{author}{McCabe, R.~J.},
  \bibinfo{author}{Beyerlein, I.~J.}, \& \bibinfo{author}{Knezevic, M.}
  (\bibinfo{year}{2015}).
\newblock \bibinfo{title}{Explicit incorporation of deformation twins into
  crystal plasticity finite element models}.
\newblock {\it \bibinfo{journal}{Computer Methods in Applied Mechanics and
  Engineering}\/},  {\it \bibinfo{volume}{295}\/}, \bibinfo{pages}{396--413}.
  \DOIprefix\doi{https://doi.org/10.1016/j.cma.2015.07.003}.
\bibitem[{Asaro \& Needleman(1985)}]{Asaro85}
\bibinfo{author}{Asaro, R.}, \& \bibinfo{author}{Needleman, A.}
  (\bibinfo{year}{1985}).
\newblock \bibinfo{title}{Overview no. 42 texture development and strain
  hardening in rate dependent polycrystals}.
\newblock {\it \bibinfo{journal}{Acta Metallurgica}\/},  {\it
  \bibinfo{volume}{33}\/}, \bibinfo{pages}{923--953}.
  \DOIprefix\doi{https://doi.org/10.1016/0001-6160(85)90188-9}.
\bibitem[{Baweja \& Joshi(2024)}]{Baweja24}
\bibinfo{author}{Baweja, S.}, \& \bibinfo{author}{Joshi, S.~P.}
  (\bibinfo{year}{2024}).
\newblock \bibinfo{title}{Microstructural effects in rate-dependent responses
  of smooth and notched magnesium bars}.
\newblock {\it \bibinfo{journal}{Mechanics of Materials}\/},  {\it
  \bibinfo{volume}{189}\/}, \bibinfo{pages}{104877}. \URLprefix
  \url{https://www.sciencedirect.com/science/article/pii/S016766362300323X}.
  \DOIprefix\doi{https://doi.org/10.1016/j.mechmat.2023.104877}.
\bibitem[{Bayraktar et~al.(2004)Bayraktar, Khalid \&
  Levaillant}]{bayraktar2004deformation}
\bibinfo{author}{Bayraktar, E.}, \bibinfo{author}{Khalid, F.}, \&
  \bibinfo{author}{Levaillant, C.} (\bibinfo{year}{2004}).
\newblock \bibinfo{title}{Deformation and fracture behaviour of high manganese
  austenitic steel}.
\newblock {\it \bibinfo{journal}{J. Mater. Process. Technol.}\/},  {\it
  \bibinfo{volume}{147}\/}, \bibinfo{pages}{145--154}.
\bibitem[{Canadinc et~al.(2007)Canadinc, Sehitoglu, Maier, Niklasch \&
  Chumlyakov}]{canadinc2007orientation}
\bibinfo{author}{Canadinc, D.}, \bibinfo{author}{Sehitoglu, H.},
  \bibinfo{author}{Maier, H.}, \bibinfo{author}{Niklasch, D.}, \&
  \bibinfo{author}{Chumlyakov, Y.} (\bibinfo{year}{2007}).
\newblock \bibinfo{title}{Orientation evolution in hadfield steel single
  crystals under combined slip and twinning}.
\newblock {\it \bibinfo{journal}{Int. J. Solid Struct.}\/},  {\it
  \bibinfo{volume}{44}\/}, \bibinfo{pages}{34--50}.
\bibitem[{Chang \& Kochmann(2015)}]{Chang15}
\bibinfo{author}{Chang, Y.}, \& \bibinfo{author}{Kochmann, D.~M.}
  (\bibinfo{year}{2015}).
\newblock \bibinfo{title}{A variational constitutive model for slip-twinning
  interactions in hcp metals: Application to single- and polycrystalline
  magnesium}.
\newblock {\it \bibinfo{journal}{International Journal of Plasticity}\/},  {\it
  \bibinfo{volume}{73}\/}, \bibinfo{pages}{39--61}.
  \DOIprefix\doi{https://doi.org/10.1016/j.ijplas.2015.03.008}.
\newblock \bibinfo{note}{Special Issue on Constitutive Modeling from
  Micro-Scale to Continuum in Honor of Prof. Frédéric Barlat}.
\bibitem[{Cheng et~al.(2018)Cheng, Shen, Mishra \& Ghosh}]{Cheng18}
\bibinfo{author}{Cheng, J.}, \bibinfo{author}{Shen, J.},
  \bibinfo{author}{Mishra, R.~K.}, \& \bibinfo{author}{Ghosh, S.}
  (\bibinfo{year}{2018}).
\newblock \bibinfo{title}{Discrete twin evolution in mg alloys using a novel
  crystal plasticity finite element model}.
\newblock {\it \bibinfo{journal}{Acta Materialia}\/},  {\it
  \bibinfo{volume}{149}\/}, \bibinfo{pages}{142--153}.
  \DOIprefix\doi{https://doi.org/10.1016/j.actamat.2018.02.032}.
\bibitem[{Cheng et~al.(2024)Cheng, Qiao, Fu, Xin, He, Huang, Chen, Wu \&
  Liu}]{Cheng24}
\bibinfo{author}{Cheng, Y.}, \bibinfo{author}{Qiao, H.}, \bibinfo{author}{Fu,
  Y.}, \bibinfo{author}{Xin, Y.}, \bibinfo{author}{He, Q.},
  \bibinfo{author}{Huang, X.}, \bibinfo{author}{Chen, G.}, \bibinfo{author}{Wu,
  P.}, \& \bibinfo{author}{Liu, Q.} (\bibinfo{year}{2024}).
\newblock \bibinfo{title}{Mechanical anisotropy induced by the competition
  between twinning and basal slip of az31 magnesium alloy under biaxial
  tension}.
\newblock {\it \bibinfo{journal}{International Journal of Plasticity}\/},  {\it
  \bibinfo{volume}{178}\/}, \bibinfo{pages}{104005}.
  \DOIprefix\doi{https://doi.org/10.1016/j.ijplas.2024.104005}.
\bibitem[{Chester et~al.(2016)Chester, Bernier, Barton, Balogh, Clausen \&
  Edmiston}]{Chester16}
\bibinfo{author}{Chester, S.~A.}, \bibinfo{author}{Bernier, J.~V.},
  \bibinfo{author}{Barton, N.~R.}, \bibinfo{author}{Balogh, L.},
  \bibinfo{author}{Clausen, B.}, \& \bibinfo{author}{Edmiston, J.~K.}
  (\bibinfo{year}{2016}).
\newblock \bibinfo{title}{Direct numerical simulation of deformation twinning
  in polycrystals}.
\newblock {\it \bibinfo{journal}{Acta Materialia}\/},  {\it
  \bibinfo{volume}{120}\/}, \bibinfo{pages}{348--363}.
  \DOIprefix\doi{https://doi.org/10.1016/j.actamat.2016.08.054}.
\bibitem[{Chin et~al.(1969)Chin, Hosford, Mendorf \& Taylor}]{Chin69}
\bibinfo{author}{Chin, G.~Y.}, \bibinfo{author}{Hosford, W.~F.},
  \bibinfo{author}{Mendorf, D.~R.}, \& \bibinfo{author}{Taylor, G.~I.}
  (\bibinfo{year}{1969}).
\newblock \bibinfo{title}{Accommodation of constrained deformation in f. c. c.
  metals by slip and twinning}.
\newblock {\it \bibinfo{journal}{Proceedings of the Royal Society of London. A.
  Mathematical and Physical Sciences}\/},  {\it \bibinfo{volume}{309}\/},
  \bibinfo{pages}{433--456}. \DOIprefix\doi{10.1098/rspa.1969.0051}.
\bibitem[{Clausen et~al.(1998)Clausen, Lorentzen \& Leffers}]{Clausen98}
\bibinfo{author}{Clausen, B.}, \bibinfo{author}{Lorentzen, T.}, \&
  \bibinfo{author}{Leffers, T.} (\bibinfo{year}{1998}).
\newblock \bibinfo{title}{Self-consistent modelling of the plastic deformation
  of f.c.c. polycrystals and its implications for diffraction measurements of
  internal stresses}.
\newblock {\it \bibinfo{journal}{Acta Materialia}\/},  {\it
  \bibinfo{volume}{46}\/}, \bibinfo{pages}{3087--3098}.
  \DOIprefix\doi{https://doi.org/10.1016/S1359-6454(98)00014-7}.
\bibitem[{Dastur \& Leslie(1981)}]{dastur1981mechanism}
\bibinfo{author}{Dastur, Y.}, \& \bibinfo{author}{Leslie, W.}
  (\bibinfo{year}{1981}).
\newblock \bibinfo{title}{Mechanism of work hardening in hadfield manganese
  steel}.
\newblock {\it \bibinfo{journal}{Metall. Trans. A}\/},  {\it
  \bibinfo{volume}{12}\/}, \bibinfo{pages}{749--759}.
\bibitem[{Ding et~al.(2024)Ding, Dong, Huang, Guo, Tian \& Yan}]{Ding24}
\bibinfo{author}{Ding, Z.}, \bibinfo{author}{Dong, L.}, \bibinfo{author}{Huang,
  Q.}, \bibinfo{author}{Guo, J.}, \bibinfo{author}{Tian, R.}, \&
  \bibinfo{author}{Yan, F.} (\bibinfo{year}{2024}).
\newblock \bibinfo{title}{Formation of intracrystalline deformation twins in
  austenitic {Mn}-steel at the early stage of plastic deformation}.
\newblock {\it \bibinfo{journal}{Journal of Materials Science \&
  Technology}\/},  {\it \bibinfo{volume}{197}\/}, \bibinfo{pages}{129--138}.
  \DOIprefix\doi{https://doi.org/10.1016/j.jmst.2024.01.049}.
\bibitem[{Dittmann \& Wulfinghoff(2023)}]{Dittmann23}
\bibinfo{author}{Dittmann, J.}, \& \bibinfo{author}{Wulfinghoff, S.}
  (\bibinfo{year}{2023}).
\newblock \bibinfo{title}{Efficient numerical strategies for an implicit volume
  fraction transfer scheme for single crystal plasticity including twinning and
  secondary plasticity on the example of magnesium}.
\newblock {\it \bibinfo{journal}{International Journal for Numerical Methods in
  Engineering}\/},  {\it \bibinfo{volume}{124}\/}, \bibinfo{pages}{4718 –
  4739}. \DOIprefix\doi{10.1002/nme.7329}.
\bibitem[{Efstathiou \& Sehitoglu(2010)}]{efstathiou2010strain}
\bibinfo{author}{Efstathiou, C.}, \& \bibinfo{author}{Sehitoglu, H.}
  (\bibinfo{year}{2010}).
\newblock \bibinfo{title}{Strain hardening and heterogeneous deformation during
  twinning in hadfield steel}.
\newblock {\it \bibinfo{journal}{Acta Mater.}\/},  {\it
  \bibinfo{volume}{58}\/}, \bibinfo{pages}{1479--1488}.
\bibitem[{{El Kadiri} \& Oppedal(2010)}]{ElKadiri10}
\bibinfo{author}{{El Kadiri}, H.}, \& \bibinfo{author}{Oppedal, A.}
  (\bibinfo{year}{2010}).
\newblock \bibinfo{title}{A crystal plasticity theory for latent hardening by
  glide twinning through dislocation transmutation and twin accommodation
  effects}.
\newblock {\it \bibinfo{journal}{Journal of the Mechanics and Physics of
  Solids}\/},  {\it \bibinfo{volume}{58}\/}, \bibinfo{pages}{613--624}.
  \DOIprefix\doi{https://doi.org/10.1016/j.jmps.2009.12.004}.
\bibitem[{Frydrych \& Kowalczyk-Gajewska(2018)}]{Frydrych18}
\bibinfo{author}{Frydrych, K.}, \& \bibinfo{author}{Kowalczyk-Gajewska, K.}
  (\bibinfo{year}{2018}).
\newblock \bibinfo{title}{Grain refinement in the equal channel angular
  pressing process: simulations using the crystal plasticity finite element
  method}.
\newblock {\it \bibinfo{journal}{Model. Simul. Mater. Sci. Eng.}\/},  {\it
  \bibinfo{volume}{26}\/}, \bibinfo{pages}{065015}.
\bibitem[{Frydrych et~al.(2021)Frydrych, Libura, Kowalewski, Maj \&
  Kowalczyk-Gajewska}]{Frydrych21}
\bibinfo{author}{Frydrych, K.}, \bibinfo{author}{Libura, T.},
  \bibinfo{author}{Kowalewski, Z.}, \bibinfo{author}{Maj, M.}, \&
  \bibinfo{author}{Kowalczyk-Gajewska, K.} (\bibinfo{year}{2021}).
\newblock \bibinfo{title}{On the role of slip, twinning and detwinning in
  magnesium alloy az31b sheet}.
\newblock {\it \bibinfo{journal}{Materials Science and Engineering: A}\/},
  {\it \bibinfo{volume}{813}\/}. \DOIprefix\doi{10.1016/j.msea.2021.141152}.
\bibitem[{Frydrych et~al.(2020)Frydrych, Maj, Urbański \&
  Kowalczyk-Gajewska}]{Frydrych20}
\bibinfo{author}{Frydrych, K.}, \bibinfo{author}{Maj, M.},
  \bibinfo{author}{Urbański, L.}, \& \bibinfo{author}{Kowalczyk-Gajewska, K.}
  (\bibinfo{year}{2020}).
\newblock \bibinfo{title}{Twinning-induced anisotropy of mechanical response of
  {AZ31B} extruded rods}.
\newblock {\it \bibinfo{journal}{Materials Science and Engineering: A}\/},
  {\it \bibinfo{volume}{771}\/}, \bibinfo{pages}{138610}.
  \DOIprefix\doi{https://doi.org/10.1016/j.msea.2019.138610}.
\bibitem[{G{\"u}rol \& Kurnaz(2020)}]{gurol2020effect}
\bibinfo{author}{G{\"u}rol, U.}, \& \bibinfo{author}{Kurnaz, S.~C.}
  (\bibinfo{year}{2020}).
\newblock \bibinfo{title}{Effect of carbon and manganese content on the
  microstructure and mechanical properties of high manganese austenitic steel}.
\newblock {\it \bibinfo{journal}{Journal of Mining and Metallurgy, Section B:
  Metallurgy}\/},  {\it \bibinfo{volume}{56}\/}, \bibinfo{pages}{171--182}.
  \DOIprefix\doi{https://doi.org/10.2298/JMMB191111009G}.
\bibitem[{Havel(2017)}]{havel2017austenitic}
\bibinfo{author}{Havel, D.} (\bibinfo{year}{2017}).
\newblock {\it \bibinfo{title}{Austenitic Manganese Steel: A Complete
  Overview}\/}.
\newblock \bibinfo{type}{Technical Report} Columbia Steel Casting Co., Inc.
\bibitem[{Havl{\'\i}{\v{c}}ek \&
  Bu{\v{s}}ov{\'a}(2012)}]{havlicek2012experience}
\bibinfo{author}{Havl{\'\i}{\v{c}}ek, P.}, \&
  \bibinfo{author}{Bu{\v{s}}ov{\'a}, K.} (\bibinfo{year}{2012}).
\newblock \bibinfo{title}{Experience with explosive hardening of railway frogs
  from hadfield steel}.
\newblock In {\it \bibinfo{booktitle}{METAL 2012 - Conference Proceedings, 21st
  International Conference on Metallurgy and Materials}\/}.
\newblock \bibinfo{publisher}{TANGER Ltd.}
\bibitem[{Herbig et~al.(2015)Herbig, Kuzmina, Haase, Marceau,
  Guti{\'e}rrez-Urrutia, Haley, Molodov, Choi \& Raabe}]{herbig2015grain}
\bibinfo{author}{Herbig, M.}, \bibinfo{author}{Kuzmina, M.},
  \bibinfo{author}{Haase, C.}, \bibinfo{author}{Marceau, R.~K.},
  \bibinfo{author}{Guti{\'e}rrez-Urrutia, I.}, \bibinfo{author}{Haley, D.},
  \bibinfo{author}{Molodov, D.~A.}, \bibinfo{author}{Choi, P.}, \&
  \bibinfo{author}{Raabe, D.} (\bibinfo{year}{2015}).
\newblock \bibinfo{title}{Grain boundary segregation in fe--mn--c
  twinning-induced plasticity steels studied by correlative electron
  backscatter diffraction and atom probe tomography}.
\newblock {\it \bibinfo{journal}{Acta Materialia}\/},  {\it
  \bibinfo{volume}{83}\/}, \bibinfo{pages}{37--47}.
  \DOIprefix\doi{https://doi.org/10.1016/j.actamat.2014.09.041}.
\bibitem[{Hill \& Rice(1972)}]{HILL1972401}
\bibinfo{author}{Hill, R.}, \& \bibinfo{author}{Rice, J.}
  (\bibinfo{year}{1972}).
\newblock \bibinfo{title}{Constitutive analysis of elastic-plastic crystals at
  arbitrary strain}.
\newblock {\it \bibinfo{journal}{Journal of the Mechanics and Physics of
  Solids}\/},  {\it \bibinfo{volume}{20}\/}, \bibinfo{pages}{401--413}.
  \DOIprefix\doi{https://doi.org/10.1016/0022-5096(72)90017-8}.
\bibitem[{Hollenweger \& Kochmann(2022)}]{Hollenweger22}
\bibinfo{author}{Hollenweger, Y.}, \& \bibinfo{author}{Kochmann, D.~M.}
  (\bibinfo{year}{2022}).
\newblock \bibinfo{title}{An efficient temperature-dependent crystal plasticity
  framework for pure magnesium with emphasis on the competition between slip
  and twinning}.
\newblock {\it \bibinfo{journal}{International Journal of Plasticity}\/},  {\it
  \bibinfo{volume}{159}\/}, \bibinfo{pages}{103448}.
  \DOIprefix\doi{https://doi.org/10.1016/j.ijplas.2022.103448}.
\bibitem[{Huang et~al.(2022)Huang, Gao, Zhou, Xin, Tang \& Elkhodary}]{Huang22}
\bibinfo{author}{Huang, C.}, \bibinfo{author}{Gao, B.}, \bibinfo{author}{Zhou,
  N.}, \bibinfo{author}{Xin, R.}, \bibinfo{author}{Tang, S.}, \&
  \bibinfo{author}{Elkhodary, K.} (\bibinfo{year}{2022}).
\newblock \bibinfo{title}{Enabling high-fidelity twin pattern prediction in
  polycrystals — a mesoscale grain boundary plasticity model.}
\newblock {\it \bibinfo{journal}{International Journal of Plasticity}\/},  {\it
  \bibinfo{volume}{148}\/}, \bibinfo{pages}{103121}.
  \DOIprefix\doi{https://doi.org/10.1016/j.ijplas.2021.103121}.
\bibitem[{Huang et~al.(2025)Huang, Lei, Yang \& Wen}]{Huang25}
\bibinfo{author}{Huang, J.}, \bibinfo{author}{Lei, M.}, \bibinfo{author}{Yang,
  G.}, \& \bibinfo{author}{Wen, B.} (\bibinfo{year}{2025}).
\newblock \bibinfo{title}{Twinning induced by asymmetric shear response}.
\newblock {\it \bibinfo{journal}{International Journal of Plasticity}\/},  {\it
  \bibinfo{volume}{185}\/}, \bibinfo{pages}{104226}.
  \DOIprefix\doi{https://doi.org/10.1016/j.ijplas.2024.104226}.
\bibitem[{Hutchinson \& Ridley(2006)}]{hutchinson2006dislocation}
\bibinfo{author}{Hutchinson, B.}, \& \bibinfo{author}{Ridley, N.}
  (\bibinfo{year}{2006}).
\newblock \bibinfo{title}{On dislocation accumulation and work hardening in
  hadfield steel}.
\newblock {\it \bibinfo{journal}{Scripta Mater.}\/},  {\it
  \bibinfo{volume}{55}\/}, \bibinfo{pages}{299--302}.
\bibitem[{Kalidindi(1998)}]{Kalidindi98}
\bibinfo{author}{Kalidindi, S.~R.} (\bibinfo{year}{1998}).
\newblock \bibinfo{title}{Incorporation of deformation twinning in crystal
  plasticity models}.
\newblock {\it \bibinfo{journal}{Journal of the Mechanics and Physics of
  Solids}\/},  {\it \bibinfo{volume}{46}\/}, \bibinfo{pages}{267 -- 290}.
  \DOIprefix\doi{https://doi.org/10.1016/S0022-5096(97)00051-3}.
\bibitem[{Kalidindi(2001)}]{Kalidindi01}
\bibinfo{author}{Kalidindi, S.~R.} (\bibinfo{year}{2001}).
\newblock \bibinfo{title}{Modeling anisotropic strain hardening and deformation
  textures in low stacking fault energy fcc metals}.
\newblock {\it \bibinfo{journal}{International Journal of Plasticity}\/},  {\it
  \bibinfo{volume}{17}\/}, \bibinfo{pages}{837 -- 860}.
  \DOIprefix\doi{https://doi.org/10.1016/S0749-6419(00)00071-1}.
\bibitem[{Karaman et~al.(2000{\natexlab{a}})Karaman, Sehitoglu, Beaudoin,
  Chumlyakov, Maier \& Tom{\'{e}}}]{Karaman00}
\bibinfo{author}{Karaman, I.}, \bibinfo{author}{Sehitoglu, H.},
  \bibinfo{author}{Beaudoin, A.}, \bibinfo{author}{Chumlyakov, Y.},
  \bibinfo{author}{Maier, H.}, \& \bibinfo{author}{Tom{\'{e}}, C.}
  (\bibinfo{year}{2000}{\natexlab{a}}).
\newblock \bibinfo{title}{Modeling the deformation behavior of {Hadfield} steel
  single and polycrystals due to twinning and slip}.
\newblock {\it \bibinfo{journal}{Acta Materialia}\/},  {\it
  \bibinfo{volume}{48}\/}, \bibinfo{pages}{2031 -- 2047}.
  \DOIprefix\doi{https://doi.org/10.1016/S1359-6454(00)00051-3}.
\bibitem[{Karaman et~al.(2001{\natexlab{a}})Karaman, Sehitoglu, Chumlyakov,
  Maier \& Kireeva}]{karaman2001extrinsic}
\bibinfo{author}{Karaman, I.}, \bibinfo{author}{Sehitoglu, H.},
  \bibinfo{author}{Chumlyakov, Y.}, \bibinfo{author}{Maier, H.}, \&
  \bibinfo{author}{Kireeva, I.} (\bibinfo{year}{2001}{\natexlab{a}}).
\newblock \bibinfo{title}{Extrinsic stacking faults and twinning in {Hadfield}
  manganese steel single crystals}.
\newblock {\it \bibinfo{journal}{Scripta Mater.}\/},  {\it
  \bibinfo{volume}{44}\/}, \bibinfo{pages}{337--343}.
\bibitem[{Karaman et~al.(2001{\natexlab{b}})Karaman, Sehitoglu, Chumlyakov,
  Maier \& Kireeva}]{Karaman01}
\bibinfo{author}{Karaman, I.}, \bibinfo{author}{Sehitoglu, H.},
  \bibinfo{author}{Chumlyakov, Y.~I.}, \bibinfo{author}{Maier, H.~J.}, \&
  \bibinfo{author}{Kireeva, I.~V.} (\bibinfo{year}{2001}{\natexlab{b}}).
\newblock \bibinfo{title}{The effect of twinning and slip on the {B}auschinger
  effect of {H}adfield steel single crystals}.
\newblock {\it \bibinfo{journal}{Metal. Mater. Trans. A}\/},  {\it
  \bibinfo{volume}{32A}\/}, \bibinfo{pages}{695--706}.
\bibitem[{Karaman et~al.(1998)Karaman, Sehitoglu, Gall \&
  Chumlyakov}]{karaman1998mechanisms}
\bibinfo{author}{Karaman, I.}, \bibinfo{author}{Sehitoglu, H.},
  \bibinfo{author}{Gall, K.}, \& \bibinfo{author}{Chumlyakov, Y.}
  (\bibinfo{year}{1998}).
\newblock \bibinfo{title}{On the deformation mechanisms in single crystal
  {Hadfield} manganese steels}.
\newblock {\it \bibinfo{journal}{Scripta Mater.}\/},  {\it
  \bibinfo{volume}{38}\/}, \bibinfo{pages}{1009--1015}.
\bibitem[{Karaman et~al.(2000{\natexlab{b}})Karaman, Sehitoglu, Gall,
  Chumlyakov \& Maier}]{karaman2000deformation}
\bibinfo{author}{Karaman, I.}, \bibinfo{author}{Sehitoglu, H.},
  \bibinfo{author}{Gall, K.}, \bibinfo{author}{Chumlyakov, Y.}, \&
  \bibinfo{author}{Maier, H.} (\bibinfo{year}{2000}{\natexlab{b}}).
\newblock \bibinfo{title}{Deformation of single crystal {Hadfield} steel by
  twinning and slip}.
\newblock {\it \bibinfo{journal}{Acta Mater.}\/},  {\it
  \bibinfo{volume}{48}\/}, \bibinfo{pages}{1345--1359}.
\bibitem[{Kaushik et~al.(2014{\natexlab{a}})Kaushik, Narasimhan \&
  Mishra}]{kaushik2014experimental}
\bibinfo{author}{Kaushik, V.}, \bibinfo{author}{Narasimhan, R.}, \&
  \bibinfo{author}{Mishra, R.} (\bibinfo{year}{2014}{\natexlab{a}}).
\newblock \bibinfo{title}{Experimental study of fracture behavior of magnesium
  single crystals}.
\newblock {\it \bibinfo{journal}{Materials Science and Engineering: A}\/},
  {\it \bibinfo{volume}{590}\/}, \bibinfo{pages}{174--185}.
\bibitem[{Kaushik et~al.(2014{\natexlab{b}})Kaushik, Narasimhan \&
  Mishra}]{kaushik2014finite}
\bibinfo{author}{Kaushik, V.}, \bibinfo{author}{Narasimhan, R.}, \&
  \bibinfo{author}{Mishra, R.~K.} (\bibinfo{year}{2014}{\natexlab{b}}).
\newblock \bibinfo{title}{Finite element simulations of notch tip fields in
  magnesium single crystals}.
\newblock {\it \bibinfo{journal}{International Journal of Fracture}\/},  {\it
  \bibinfo{volume}{189}\/}, \bibinfo{pages}{195--216}.
\bibitem[{Korelc(2002)}]{Korelc02}
\bibinfo{author}{Korelc, J.} (\bibinfo{year}{2002}).
\newblock \bibinfo{title}{{Multi-language and multi-environment generation of
  nonlinear finite element codes}}.
\newblock {\it \bibinfo{journal}{Eng. Comput.}\/},  {\it
  \bibinfo{volume}{18}\/}, \bibinfo{pages}{312--327}.
\bibitem[{Korelc \& Stupkiewicz(2014)}]{Korelc14}
\bibinfo{author}{Korelc, J.}, \& \bibinfo{author}{Stupkiewicz, S.}
  (\bibinfo{year}{2014}).
\newblock \bibinfo{title}{Closed-form matrix exponential and its application in
  finite-strain plasticity}.
\newblock {\it \bibinfo{journal}{International Journal for Numerical Methods in
  Engineering}\/},  {\it \bibinfo{volume}{98}\/}, \bibinfo{pages}{960 – 987}.
  \DOIprefix\doi{10.1002/nme.4653}.
\bibitem[{Kowalczyk-Gajewska(2010)}]{Kowalczyk10}
\bibinfo{author}{Kowalczyk-Gajewska, K.} (\bibinfo{year}{2010}).
\newblock \bibinfo{title}{Modelling of texture evolution in metals accounting
  for lattice reorientation due to twinning}.
\newblock {\it \bibinfo{journal}{European Journal of Mechanics - A/Solids}\/},
  {\it \bibinfo{volume}{29}\/}, \bibinfo{pages}{28 -- 41}.
  \DOIprefix\doi{https://doi.org/10.1016/j.euromechsol.2009.07.002}.
\bibitem[{Kowalczyk-Gajewska(2011)}]{Kowalczyk11}
\bibinfo{author}{Kowalczyk-Gajewska, K.} (\bibinfo{year}{2011}).
\newblock \bibinfo{title}{Micromechanical modelling of metals and alloys of
  high specific strength}.
\newblock {\it \bibinfo{journal}{IFTR Reports 1/2011}\/},  (pp.
  \bibinfo{pages}{1--299}).
\bibitem[{Kowalczyk-Gajewska(2013)}]{Kowalczyk13}
\bibinfo{author}{Kowalczyk-Gajewska, K.} (\bibinfo{year}{2013}).
\newblock \bibinfo{title}{Crystal plasticity models accounting for twinning}.
\newblock {\it \bibinfo{journal}{Computer Methods in Materials Science}\/},
  {\it \bibinfo{volume}{13}\/}, \bibinfo{pages}{436--451}.
\bibitem[{Kucharski et~al.(2014)Kucharski, Stupkiewicz \& Petryk}]{Kucharski14}
\bibinfo{author}{Kucharski, S.}, \bibinfo{author}{Stupkiewicz, S.}, \&
  \bibinfo{author}{Petryk, H.} (\bibinfo{year}{2014}).
\newblock \bibinfo{title}{Surface pile-up patterns in indentation testing of cu
  single crystals}.
\newblock {\it \bibinfo{journal}{Exper. Mech.}\/},  {\it
  \bibinfo{volume}{54}\/}, \bibinfo{pages}{957--969}.
  \DOIprefix\doi{https://doi.org/10.1007/s11340-014-9883-1}.
\bibitem[{Lebensohn \& Tom{\'{e}}(1993)}]{Lebensohn93}
\bibinfo{author}{Lebensohn, R.}, \& \bibinfo{author}{Tom{\'{e}}, C.}
  (\bibinfo{year}{1993}).
\newblock \bibinfo{title}{A self-consistent anisotropic approach for the
  simulation of plastic deformation and texture development of polycrystals:
  {A}pplication to zirconium alloys}.
\newblock {\it \bibinfo{journal}{Acta Metallurgica et Materialia}\/},  {\it
  \bibinfo{volume}{41}\/}, \bibinfo{pages}{2611 -- 2624}.
  \DOIprefix\doi{https://doi.org/10.1016/0956-7151(93)90130-K}.
\bibitem[{Lindroos et~al.(2018)Lindroos, Cailletaud, Laukkanen \&
  Kuokkala}]{Lindroos18}
\bibinfo{author}{Lindroos, M.}, \bibinfo{author}{Cailletaud, G.},
  \bibinfo{author}{Laukkanen, A.}, \& \bibinfo{author}{Kuokkala, V.-T.}
  (\bibinfo{year}{2018}).
\newblock \bibinfo{title}{Crystal plasticity modeling and characterization of
  the deformation twinning and strain hardening in hadfield steels}.
\newblock {\it \bibinfo{journal}{Materials Science and Engineering: A}\/},
  {\it \bibinfo{volume}{720}\/}, \bibinfo{pages}{145--159}.
  \DOIprefix\doi{https://doi.org/10.1016/j.msea.2018.02.028}.
\bibitem[{Liu et~al.(2023)Liu, Roters \& Raabe}]{LiuRotersRaabe23}
\bibinfo{author}{Liu, C.}, \bibinfo{author}{Roters, F.}, \&
  \bibinfo{author}{Raabe, D.} (\bibinfo{year}{2023}).
\newblock \bibinfo{title}{Finite strain crystal plasticity-phase field modeling
  of twin, dislocation, and grain boundary interaction in hexagonal materials}.
\newblock {\it \bibinfo{journal}{Acta Materialia}\/},  {\it
  \bibinfo{volume}{242}\/}, \bibinfo{pages}{118444}.
  \DOIprefix\doi{https://doi.org/10.1016/j.actamat.2022.118444}.
\bibitem[{Liu \& Wei(2014)}]{LiuWei14}
\bibinfo{author}{Liu, Y.}, \& \bibinfo{author}{Wei, Y.} (\bibinfo{year}{2014}).
\newblock \bibinfo{title}{A polycrystal based numerical investigation on the
  temperature dependence of slip resistance and texture evolution in magnesium
  alloy az31b}.
\newblock {\it \bibinfo{journal}{International Journal of Plasticity}\/},  {\it
  \bibinfo{volume}{55}\/}, \bibinfo{pages}{80--93}.
  \DOIprefix\doi{https://doi.org/10.1016/j.ijplas.2013.09.011}.
\bibitem[{M{\`{a}}nik et~al.(2022)M{\`{a}}nik, Asadkandi \& Holmedal}]{Manik22}
\bibinfo{author}{M{\`{a}}nik, T.}, \bibinfo{author}{Asadkandi, H.}, \&
  \bibinfo{author}{Holmedal, B.} (\bibinfo{year}{2022}).
\newblock \bibinfo{title}{A robust algorithm for rate-independent crystal
  plasticity}.
\newblock {\it \bibinfo{journal}{Computer Methods in Applied Mechanics and
  Engineering}\/},  {\it \bibinfo{volume}{393}\/}.
  \DOIprefix\doi{10.1016/j.cma.2022.114831}.
\bibitem[{Meyers et~al.(2001)Meyers, Vöhringer \& Lubarda}]{Meyers01}
\bibinfo{author}{Meyers, M.}, \bibinfo{author}{Vöhringer, O.}, \&
  \bibinfo{author}{Lubarda, V.} (\bibinfo{year}{2001}).
\newblock \bibinfo{title}{The onset of twinning in metals: a constitutive
  description}.
\newblock {\it \bibinfo{journal}{Acta Materialia}\/},  {\it
  \bibinfo{volume}{49}\/}, \bibinfo{pages}{4025--4039}. \URLprefix
  \url{https://www.sciencedirect.com/science/article/pii/S1359645401003007}.
  \DOIprefix\doi{https://doi.org/10.1016/S1359-6454(01)00300-7}.
\bibitem[{Mo et~al.(2022)Mo, Liu, Mao, Shen \& Wang}]{Mo22}
\bibinfo{author}{Mo, H.}, \bibinfo{author}{Liu, G.}, \bibinfo{author}{Mao, Y.},
  \bibinfo{author}{Shen, Y.}, \& \bibinfo{author}{Wang, J.}
  (\bibinfo{year}{2022}).
\newblock \bibinfo{title}{Dual-interface model for twinning in the coupled
  crystal plasticity finite element – phase field method}.
\newblock {\it \bibinfo{journal}{International Journal of Plasticity}\/},  {\it
  \bibinfo{volume}{158}\/}, \bibinfo{pages}{103441}.
  \DOIprefix\doi{https://doi.org/10.1016/j.ijplas.2022.103441}.
\bibitem[{Okechukwu et~al.(2018)Okechukwu, Dahunsi, Oke, Oladele \&
  Dauda}]{okechukwu2018development}
\bibinfo{author}{Okechukwu, C.}, \bibinfo{author}{Dahunsi, O.~A.},
  \bibinfo{author}{Oke, P.~K.}, \bibinfo{author}{Oladele, I.~O.}, \&
  \bibinfo{author}{Dauda, M.} (\bibinfo{year}{2018}).
\newblock \bibinfo{title}{Development of hardfaced crusher jaws using
  ferro-alloy hardfacing inserts and low carbon steel substrate}.
\newblock {\it \bibinfo{journal}{J. Tribol}\/},  {\it \bibinfo{volume}{18}\/},
  \bibinfo{pages}{20--39}.
\bibitem[{Prasad et~al.(2017)Prasad, Narasimhan \& Suwas}]{Subrahmanya17}
\bibinfo{author}{Prasad, N.~S.}, \bibinfo{author}{Narasimhan, R.}, \&
  \bibinfo{author}{Suwas, S.} (\bibinfo{year}{2017}).
\newblock \bibinfo{title}{Effects of lattice orientation and crack tip
  constraint on ductile fracture initiation in mg single crystals}.
\newblock {\it \bibinfo{journal}{International Journal of Plasticity}\/},  {\it
  \bibinfo{volume}{97}\/}, \bibinfo{pages}{222--245}. \URLprefix
  \url{https://www.sciencedirect.com/science/article/pii/S0749641917302085}.
  \DOIprefix\doi{https://doi.org/10.1016/j.ijplas.2017.06.004}.
\bibitem[{Proust et~al.(2007)Proust, Tom{\'{e}} \& Kaschner}]{Proust07}
\bibinfo{author}{Proust, G.}, \bibinfo{author}{Tom{\'{e}}, C.}, \&
  \bibinfo{author}{Kaschner, G.} (\bibinfo{year}{2007}).
\newblock \bibinfo{title}{Modeling texture, twinning and hardening evolution
  during deformation of hexagonal materials}.
\newblock {\it \bibinfo{journal}{Acta Materialia}\/},  {\it
  \bibinfo{volume}{55}\/}, \bibinfo{pages}{2137 -- 2148}.
  \DOIprefix\doi{https://doi.org/10.1016/j.actamat.2006.11.017}.
\bibitem[{Qiao et~al.(2016)Qiao, Barnett \& Wu}]{Qiao16}
\bibinfo{author}{Qiao, H.}, \bibinfo{author}{Barnett, M.}, \&
  \bibinfo{author}{Wu, P.} (\bibinfo{year}{2016}).
\newblock \bibinfo{title}{Modeling of twin formation, propagation and growth in
  a mg single crystal based on crystal plasticity finite element method}.
\newblock {\it \bibinfo{journal}{International Journal of Plasticity}\/},  {\it
  \bibinfo{volume}{86}\/}, \bibinfo{pages}{70--92}.
  \DOIprefix\doi{https://doi.org/10.1016/j.ijplas.2016.08.002}.
\bibitem[{Qiao et~al.(2017)Qiao, Guo, Hong \& Wu}]{Qiao17}
\bibinfo{author}{Qiao, H.}, \bibinfo{author}{Guo, X.}, \bibinfo{author}{Hong,
  S.}, \& \bibinfo{author}{Wu, P.} (\bibinfo{year}{2017}).
\newblock \bibinfo{title}{Modeling of {10-12}-{10-12} secondary twinning in
  pre-compressed {M}g alloy {AZ31}}.
\newblock {\it \bibinfo{journal}{Journal of Alloys and Compounds}\/},  {\it
  \bibinfo{volume}{725}\/}, \bibinfo{pages}{96--107}. \URLprefix
  \url{https://www.sciencedirect.com/science/article/pii/S0925838817324982}.
  \DOIprefix\doi{https://doi.org/10.1016/j.jallcom.2017.07.133}.
\bibitem[{Raghavan et~al.(1969)Raghavan, Sastri \&
  Marcinkowski}]{raghavan1969nature}
\bibinfo{author}{Raghavan, K.}, \bibinfo{author}{Sastri, A.}, \&
  \bibinfo{author}{Marcinkowski, M.} (\bibinfo{year}{1969}).
\newblock \bibinfo{title}{Nature of the work-hardening behavior in hadfields
  manganese steel}.
\newblock {\it \bibinfo{journal}{Trans. Met. Soc. AIME}\/},  {\it
  \bibinfo{volume}{245}\/}, \bibinfo{pages}{1569--1575}.
\bibitem[{Rezaee-Hajidehi et~al.(2022)Rezaee-Hajidehi, Sadowski \&
  Stupkiewicz}]{RezaeeHajidehi22}
\bibinfo{author}{Rezaee-Hajidehi, M.}, \bibinfo{author}{Sadowski, P.}, \&
  \bibinfo{author}{Stupkiewicz, S.} (\bibinfo{year}{2022}).
\newblock \bibinfo{title}{Deformation twinning as a displacive transformation:
  Finite-strain phase-field model of coupled twinning and crystal plasticity}.
\newblock {\it \bibinfo{journal}{Journal of the Mechanics and Physics of
  Solids}\/},  {\it \bibinfo{volume}{163}\/}, \bibinfo{pages}{104855}.
  \DOIprefix\doi{https://doi.org/10.1016/j.jmps.2022.104855}.
\bibitem[{Sahoo et~al.(2020)Sahoo, Biswas, Toth, Gautam \& Beausir}]{Sahoo20}
\bibinfo{author}{Sahoo, S.~K.}, \bibinfo{author}{Biswas, S.},
  \bibinfo{author}{Toth, L.~S.}, \bibinfo{author}{Gautam, P.}, \&
  \bibinfo{author}{Beausir, B.} (\bibinfo{year}{2020}).
\newblock \bibinfo{title}{Strain hardening, twinning and texture evolution in
  magnesium alloy using the all twin variant polycrystal modelling approach}.
\newblock {\it \bibinfo{journal}{International Journal of Plasticity}\/},  {\it
  \bibinfo{volume}{128}\/}, \bibinfo{pages}{102660}.
  \DOIprefix\doi{https://doi.org/10.1016/j.ijplas.2020.102660}.
\bibitem[{Sahoo et~al.(2019)Sahoo, Toth \& Biswas}]{Sahoo19}
\bibinfo{author}{Sahoo, S.~K.}, \bibinfo{author}{Toth, L.~S.}, \&
  \bibinfo{author}{Biswas, S.} (\bibinfo{year}{2019}).
\newblock \bibinfo{title}{An analytical model to predict strain-hardening
  behaviour and twin volume fraction in a profoundly twinning magnesium alloy}.
\newblock {\it \bibinfo{journal}{International Journal of Plasticity}\/},  {\it
  \bibinfo{volume}{119}\/}, \bibinfo{pages}{273 -- 290}.
  \DOIprefix\doi{https://doi.org/10.1016/j.ijplas.2019.04.007}.
\bibitem[{Sahu et~al.(2024)Sahu, Chakraborty, Yadava \& Gurao}]{Sahu24}
\bibinfo{author}{Sahu, V.~K.}, \bibinfo{author}{Chakraborty, P.},
  \bibinfo{author}{Yadava, M.}, \& \bibinfo{author}{Gurao, N.~P.}
  (\bibinfo{year}{2024}).
\newblock \bibinfo{title}{Micro-mechanisms of anisotropic deformation in the
  presence of notch in commercially pure titanium: An in-situ study with cpfem
  simulations}.
\newblock {\it \bibinfo{journal}{International Journal of Plasticity}\/},  {\it
  \bibinfo{volume}{177}\/}. \DOIprefix\doi{10.1016/j.ijplas.2024.103985}.
\bibitem[{Selvarajou et~al.(2016)Selvarajou, Kondori, Benzerga \&
  Joshi}]{Selvarajou16}
\bibinfo{author}{Selvarajou, B.}, \bibinfo{author}{Kondori, B.},
  \bibinfo{author}{Benzerga, A.~A.}, \& \bibinfo{author}{Joshi, S.~P.}
  (\bibinfo{year}{2016}).
\newblock \bibinfo{title}{On plastic flow in notched hexagonal close packed
  single crystals}.
\newblock {\it \bibinfo{journal}{Journal of the Mechanics and Physics of
  Solids}\/},  {\it \bibinfo{volume}{94}\/}, \bibinfo{pages}{273--297}.
  \DOIprefix\doi{https://doi.org/10.1016/j.jmps.2016.04.023}.
\bibitem[{Shen et~al.(2024)Shen, Xie, Wu, Luo, haifeng Ye \& Chen}]{Shen24}
\bibinfo{author}{Shen, S.}, \bibinfo{author}{Xie, P.}, \bibinfo{author}{Wu,
  C.}, \bibinfo{author}{Luo, J.}, \bibinfo{author}{haifeng Ye}, \&
  \bibinfo{author}{Chen, J.} (\bibinfo{year}{2024}).
\newblock \bibinfo{title}{Cross-slip of extended dislocations and secondary
  deformation twinning in a high-mn twip steel}.
\newblock {\it \bibinfo{journal}{International Journal of Plasticity}\/},  {\it
  \bibinfo{volume}{175}\/}, \bibinfo{pages}{103922}.
  \DOIprefix\doi{https://doi.org/10.1016/j.ijplas.2024.103922}.
\bibitem[{Simo \& Hughes(1998)}]{Simo98}
\bibinfo{author}{Simo, J.~C.}, \& \bibinfo{author}{Hughes, T. J.~R.}
  (\bibinfo{year}{1998}).
\newblock {\it \bibinfo{title}{Computational Inelasticity}\/}.
\newblock \bibinfo{publisher}{Springer}.
\bibitem[{Srivastava \& Needleman(2015)}]{SRIVASTAVA201510}
\bibinfo{author}{Srivastava, A.}, \& \bibinfo{author}{Needleman, A.}
  (\bibinfo{year}{2015}).
\newblock \bibinfo{title}{Effect of crystal orientation on porosity evolution
  in a creeping single crystal}.
\newblock {\it \bibinfo{journal}{Mechanics of Materials}\/},  {\it
  \bibinfo{volume}{90}\/}, \bibinfo{pages}{10--29}. \URLprefix
  \url{https://www.sciencedirect.com/science/article/pii/S0167663615000228}.
  \DOIprefix\doi{https://doi.org/10.1016/j.mechmat.2015.01.015}.
\newblock \bibinfo{note}{Proceedings of the IUTAM Symposium on Micromechanics
  of Defects in Solids}.
\bibitem[{Staroselsky \& Anand(1998)}]{Staroselsky98}
\bibinfo{author}{Staroselsky, A.}, \& \bibinfo{author}{Anand, L.}
  (\bibinfo{year}{1998}).
\newblock \bibinfo{title}{Inelastic deformation of polycrystalline face
  centered cubic materials by slip and twinning}.
\newblock {\it \bibinfo{journal}{Journal of the Mechanics and Physics of
  Solids}\/},  {\it \bibinfo{volume}{46}\/}, \bibinfo{pages}{671 -- 696}.
  \DOIprefix\doi{https://doi.org/10.1016/S0022-5096(97)00071-9}.
\bibitem[{Staroselsky \& Anand(2003)}]{Staroselsky03}
\bibinfo{author}{Staroselsky, A.}, \& \bibinfo{author}{Anand, L.}
  (\bibinfo{year}{2003}).
\newblock \bibinfo{title}{A constitutive model for hcp materials deforming by
  slip and twinning: application to magnesium alloy {AZ31B}}.
\newblock {\it \bibinfo{journal}{International Journal of Plasticity}\/},  {\it
  \bibinfo{volume}{19}\/}, \bibinfo{pages}{1843 -- 1864}.
  \DOIprefix\doi{https://doi.org/10.1016/S0749-6419(03)00039-1}.
\bibitem[{Subramanyam et~al.(1990)Subramanyam, Swansiger \&
  Avery}]{subramanyam1990austenitic}
\bibinfo{author}{Subramanyam, D.}, \bibinfo{author}{Swansiger, A.}, \&
  \bibinfo{author}{Avery, H.} (\bibinfo{year}{1990}).
\newblock \bibinfo{title}{Austenitic manganese steels}.
\newblock In \bibinfo{editor}{{ASM Handbook Committee}} (Ed.), {\it
  \bibinfo{booktitle}{Properties and Selection: Irons, Steels, and
  High-Performance Alloys}\/} (pp. \bibinfo{pages}{822--840}).
\newblock \bibinfo{address}{Materials Park, OH}: \bibinfo{publisher}{ASM
  International} volume~\bibinfo{volume}{1}.
\bibitem[{Szczerba et~al.(2004)Szczerba, Bajor \& Tokarski}]{Szczerba04}
\bibinfo{author}{Szczerba, M.}, \bibinfo{author}{Bajor, T.}, \&
  \bibinfo{author}{Tokarski, T.} (\bibinfo{year}{2004}).
\newblock \bibinfo{title}{Is there a critical resolved shear stress for
  twinning in face-centred cubic crystals?}
\newblock {\it \bibinfo{journal}{Philosophical Magazine}\/},  {\it
  \bibinfo{volume}{84}\/}, \bibinfo{pages}{481--502}.
  \DOIprefix\doi{10.1080/14786430310001612175}.
\bibitem[{Tom{\'{e}} et~al.(1991)Tom{\'{e}}, Lebensohn \& Kocks}]{Tome91}
\bibinfo{author}{Tom{\'{e}}, C.}, \bibinfo{author}{Lebensohn, R.}, \&
  \bibinfo{author}{Kocks, U.} (\bibinfo{year}{1991}).
\newblock \bibinfo{title}{A model for texture development dominated by
  deformation twinning: {A}pplication to zirconium alloys}.
\newblock {\it \bibinfo{journal}{Acta Metallurgica et Materialia}\/},  {\it
  \bibinfo{volume}{39}\/}, \bibinfo{pages}{2667 -- 2680}.
  \DOIprefix\doi{https://doi.org/10.1016/0956-7151(91)90083-D}.
\bibitem[{Wang et~al.(2013)Wang, Wu, Wang \& Tome}]{Wang13}
\bibinfo{author}{Wang, H.}, \bibinfo{author}{Wu, P.}, \bibinfo{author}{Wang,
  J.}, \& \bibinfo{author}{Tome, C.} (\bibinfo{year}{2013}).
\newblock \bibinfo{title}{A crystal plasticity model for hexagonal close packed
  (hcp) crystals including twinning and de-twinning mechanisms}.
\newblock {\it \bibinfo{journal}{International Journal of Plasticity}\/},  {\it
  \bibinfo{volume}{49}\/}, \bibinfo{pages}{36--52}.
  \DOIprefix\doi{https://doi.org/10.1016/j.ijplas.2013.02.016}.
\bibitem[{Wong et~al.(2016)Wong, Madivala, Prahl, Roters \& Raabe}]{Wong16}
\bibinfo{author}{Wong, S.~L.}, \bibinfo{author}{Madivala, M.},
  \bibinfo{author}{Prahl, U.}, \bibinfo{author}{Roters, F.}, \&
  \bibinfo{author}{Raabe, D.} (\bibinfo{year}{2016}).
\newblock \bibinfo{title}{A crystal plasticity model for twinning- and
  transformation-induced plasticity}.
\newblock {\it \bibinfo{journal}{Acta Materialia}\/},  {\it
  \bibinfo{volume}{118}\/}, \bibinfo{pages}{140--151}.
  \DOIprefix\doi{https://doi.org/10.1016/j.actamat.2016.07.032}.
\bibitem[{Yaghoobi et~al.(2020)Yaghoobi, Allison \&
  Sundararaghavan}]{Yaghoobi20}
\bibinfo{author}{Yaghoobi, M.}, \bibinfo{author}{Allison, J.~E.}, \&
  \bibinfo{author}{Sundararaghavan, V.} (\bibinfo{year}{2020}).
\newblock \bibinfo{title}{Multiscale modeling of twinning and detwinning
  behavior of hcp polycrystals}.
\newblock {\it \bibinfo{journal}{International Journal of Plasticity}\/},  {\it
  \bibinfo{volume}{127}\/}, \bibinfo{pages}{102653}.
  \DOIprefix\doi{https://doi.org/10.1016/j.ijplas.2019.102653}.
\bibitem[{Yaghoobi et~al.(2022)Yaghoobi, Chen, Murphy-Leonard, Sundararaghavan,
  Daly \& Allison}]{Yaghoobi22}
\bibinfo{author}{Yaghoobi, M.}, \bibinfo{author}{Chen, Z.},
  \bibinfo{author}{Murphy-Leonard, A.~D.}, \bibinfo{author}{Sundararaghavan,
  V.}, \bibinfo{author}{Daly, S.}, \& \bibinfo{author}{Allison, J.~E.}
  (\bibinfo{year}{2022}).
\newblock \bibinfo{title}{Deformation twinning and detwinning in extruded
  {Mg-4Al}: In-situ experiment and crystal plasticity simulation}.
\newblock {\it \bibinfo{journal}{International Journal of Plasticity}\/},  {\it
  \bibinfo{volume}{155}\/}, \bibinfo{pages}{103345}.
  \DOIprefix\doi{https://doi.org/10.1016/j.ijplas.2022.103345}.
\bibitem[{Zhang \& Joshi(2012)}]{Zhang12}
\bibinfo{author}{Zhang, J.}, \& \bibinfo{author}{Joshi, S.~P.}
  (\bibinfo{year}{2012}).
\newblock \bibinfo{title}{Phenomenological crystal plasticity modeling and
  detailed micromechanical investigations of pure magnesium}.
\newblock {\it \bibinfo{journal}{J. Mech. Phys. Solids}\/},  {\it
  \bibinfo{volume}{60}\/}, \bibinfo{pages}{945--972}.
\bibitem[{Zheng et~al.(2025)Zheng, Karaman \&
  Srivastava}]{zheng2025interaction}
\bibinfo{author}{Zheng, X.}, \bibinfo{author}{Karaman, I.}, \&
  \bibinfo{author}{Srivastava, A.} (\bibinfo{year}{2025}).
\newblock \bibinfo{title}{Interaction of deformation twinning and crack growth
  in single crystals of an austenitic manganese steel}.
\newblock {\it \bibinfo{journal}{Materials Science and Engineering: A}\/},
  {\it \bibinfo{volume}{924}\/}, \bibinfo{pages}{147779}.
  \DOIprefix\doi{https://doi.org/10.1016/j.msea.2024.147779}.

\end{thebibliography}
\end{document}